\documentclass[11pt]{article}

\usepackage[margin=1in,footskip=0.25in]{geometry}

\usepackage[pagewise]{lineno}

\usepackage{amsmath, amsfonts, mathtools, bm, siunitx, textcomp}

\usepackage{graphicx}
\usepackage{caption}
\usepackage{subcaption}  

\usepackage[capitalise]{cleveref}
\usepackage{xurl}
\usepackage{xr}

\usepackage{multirow}
\usepackage{color}
\usepackage{csquotes}
\usepackage{cancel}
\usepackage[normalem]{ulem}  
\usepackage{placeins}
\usepackage{comment}
\usepackage{csvsimple}
\usepackage{breqn}  
\usepackage{chemformula}  
 \usepackage{booktabs}
 \usepackage{makecell} 
 \usepackage{booktabs}
 \usepackage{tikz}
\usetikzlibrary{positioning, arrows.meta, shapes.geometric}

\newcommand{\eref}[1]{Eq.~(\ref{#1})}
\newcommand{\fref}[1]{Fig.~\ref{#1}}

\PassOptionsToPackage{normalem}{ulem}
\usepackage{ulem}

\providecolor{added}{rgb}{0,0,1}
\providecolor{deleted}{rgb}{1,0,0}

\makeatletter
\newcommand*{\addFileDependency}[1]{
  \typeout{(#1)}
  \@addtofilelist{#1}
  \IfFileExists{#1}{}{\typeout{No file #1.}}
}
\makeatother

\begin{document}

\title{Variational Quantum Latent Encoding for Topology Optimization}
\author{
  Alireza Tabarraei \\
  \small Department of Mechanical Engineering and Engineering Science, \\
  \small The University of North Carolina at Charlotte, Charlotte, NC 28223, USA \\
  \small and \\
  \small School of Data Science, The University of North Carolina at Charlotte, Charlotte, NC 28223, USA \\
  \texttt{atabarra@charlotte.edu}
}

\date{}  

\maketitle

\begin{abstract}
In this paper, a variational framework for structural topology optimization is developed, integrating quantum and classical latent encoding strategies within a coordinate-based neural decoding architecture.  In this approach, a low-dimensional latent vector—generated either by a variational quantum circuit or sampled from a Gaussian distribution—is mapped to a higher-dimensional latent space via a learnable projection layer. This enriched representation is then decoded into a high-resolution material distribution using a neural network that takes both the latent vector and Fourier-mapped spatial coordinates as input. The optimization is performed directly on the latent parameters, guided solely by physics-based objectives such as compliance minimization and volume constraints evaluated through finite element analysis, without requiring any precomputed datasets or supervised training.
Quantum latent vectors are constructed from the expectation values of Pauli observables measured on parameterized quantum circuits, providing a structured and entangled encoding of information. The classical baseline uses Gaussian-sampled latent vectors projected in the same manner. The proposed variational formulation enables the generation of diverse and physically valid topologies by exploring the latent space through sampling or perturbation, in contrast to traditional optimization methods that yield a single deterministic solution. 
Numerical experiments show that both classical and quantum encodings produce high-quality structural designs. However, quantum encodings demonstrate advantages in several benchmark cases in terms of compliance and design diversity. These results highlight the potential of quantum circuits as an effective and scalable tool for physics-constrained topology optimization and suggest promising directions for applying near-term quantum hardware in structural design.
\end{abstract}

\vspace{1em}


\section{Introduction}\label{sec1}
Topology optimization (TO) is a powerful computational framework for generating high-performance structural designs by optimally distributing material within a predefined design domain. It has been applied across a broad range of engineering disciplines, including aerospace~\cite{zhu2016topology}, mechanical~\cite{shishir2024multi, bhuiyan2025graph}, civil~\cite{mei2021structural}, and biomedical engineering~\cite{wu2021advances}, where objectives such as weight reduction, structural efficiency, and functional performance are critical~\cite{bendsoe2003topology}. Classical methods, particularly the solid isotropic material with penalization (SIMP)\cite{sigmund2001} and level-set approaches\cite{wang2003levelset}, were initially developed for compliance minimization~\cite{bendsoe2003topology,sigmund2001}, but have since been extended to a broad range of objectives including stress constraints~\cite{duysinx1998stress}, frequency optimization~\cite{yoon2010structural}, and multi-physics formulations~\cite{alexandersen2016multiphysics}. 

Despite their success, traditional topology optimization (TO) methods present several limitations. They require direct manipulation of high-dimensional design spaces, where each design variable corresponds to a discrete element in the mesh. This results in a large number of optimization variables that scale with mesh resolution, increasing computational cost and often leading to numerical issues such as checkerboarding and mesh dependency \cite{sigmund2013review}. In addition, the non-convexity of the design space makes these methods prone to convergence to suboptimal local minima. A further drawback is that these methods are inherently deterministic. For a fixed set of boundary conditions and optimization parameters, they typically produce only one solution. This lack of diversity limits the ability to explore alternative high-performing configurations, which can be especially important during early-stage conceptual design when evaluating multiple feasible options supports better decision-making and creative exploration.

To address these challenges, recent studies have proposed generative modeling approaches that reparameterize the design space through a lower-dimensional latent vector, which is decoded into a full-resolution material distribution using neural networks \cite{sosnovik2019neural,zhang2019deep,nobari2024nito}. These models reduce the optimization dimensionality and introduce structure into the design space, facilitating smoother and more efficient convergence.
Notably, several methods have explored the use of generative adversarial networks (GANs) \cite{nie2021topologygan, wang2023generative} and, more recently, diffusion-based models \cite{maze2023diffusion} to generate diverse high-quality topologies under varying constraints and loading conditions.
However, most of these generative models are based on supervised learning and require large datasets of optimized topologies for training \cite{zhang2019deep}. Models trained via autoencoders, GANs, or variational autoencoders (VAEs) may struggle to generalize beyond the training distribution and can be constrained by dataset biases or architectural limitations. Moreover, generating sufficiently diverse and high-fidelity datasets is computationally expensive, particularly for multi-physics problems or nonlinear material behaviors.

An alternative direction has emerged in the form of self-supervised or direct optimization methods, in which the network parameters are optimized from scratch for each problem using physics-based objectives such as compliance \cite{chandrasekhar2021tounn}. Neural field representations, where a neural network maps spatial coordinates to scalar densities, have gained popularity in this setting \cite{nobari2024nito}. While originally developed for computer graphics and neural rendering, they are now increasingly applied to design problems due to their ability to represent high-frequency geometry and complex structural features.

Parallel to these developments, quantum computing has introduced new opportunities for compact and expressive design parameterizations through variational quantum circuits (VQCs) \cite{cerezo2021variational,mcclean2016theory}. VQCs encode trainable quantum states via parameterized quantum gates and produce structured latent vectors by measuring expectation values of observables \cite{benedetti2019generative,romero2017quantum}. These quantum latent vectors are inherently bounded, nonlinear, and low-dimensional due to properties such as unitarity and entanglement~\cite{mitarai2018quantum,abbas2021power,grant2018hierarchical}, making them well-suited for representing global design priors in hybrid quantum-classical workflows. Unlike classical latent vectors, which are typically drawn from unstructured Gaussian distributions, quantum latent vectors provide access to a rich and structured function space that may facilitate exploration and convergence. This added structure could offer an advantage especially under dimensionality constraints, where the number of optimization variables must be kept small due to hardware or memory limitations.

Although VQCs have primarily been explored in machine learning tasks, such as quantum classifiers, GANs, and autoencoders~\cite{schuld2020circuit,romero2017quantum,khoshaman2018quantumgan}, their application to structural design and topology optimization remains nascent. Prior work has focused on reproducing classical data distributions or solving small-scale quantum chemistry benchmarks~\cite{peruzzo2014variational,endo2021hybrid}, but their integration into continuous, spatially resolved optimization frameworks has not been extensively investigated.

In this work, we propose a variational topology optimization framework that incorporates VQCs to generate low-dimensional latent vectors, which are then mapped to high-resolution material distributions through a coordinate-based neural decoder. The decoder receives both the latent vector and Fourier-mapped spatial coordinates as input, enabling it to capture fine-scale geometry and global design structure. To address the limited expressiveness imposed by small qubit counts, we introduce a learnable projection layer that transforms the quantum output into a higher-dimensional latent space compatible with the decoder. For comparison, we also implement a classical baseline where the latent vector is sampled from a standard Gaussian distribution and similarly projected to the decoder’s input space. This parallel formulation enables a direct evaluation of classical and quantum encodings within a unified, self-supervised topology optimization pipeline.

Optimization is performed end-to-end using automatic differentiation and physics-based loss functions computed via finite element analysis. This eliminates the need for traditional adjoint sensitivity analysis~\cite{bendsoe2003topology,sigmund2013review}, thereby streamlining implementation and enabling direct application to multiphysics and geometrically complex design problems. The framework is entirely self-supervised and does not rely on training datasets or generative priors.
Notably, the proposed method is inherently stochastic; different initializations of the latent vector, whether sampled from a classical Gaussian distribution or produced by a quantum circuit, can lead to distinct final topologies. This variability naturally promotes exploration of the design space and facilitates the generation of multiple diverse, high-performing solutions. Unlike conventional deterministic topology optimization methods that yield only one solution per run, the proposed framework supports scalable exploration of diverse design candidates, making it well-suited for structural optimization.

\section{Topology Optimization Formulation}

Topology optimization is a computational framework for determining the optimal material distribution within a design domain to meet performance objectives such as maximizing stiffness or minimizing compliance, subject to constraints like material volume. Unlike traditional shape or size optimization, TO allows both geometry and topology to evolve freely, enabling the discovery of complex and often non-intuitive design topologies. 
A widely studied formulation is compliance minimization, where the goal is to design the stiffest possible structure under given loads and boundary conditions while using only a limited amount of material. The design domain is discretized using the finite element method (FEM), and each element \( e \) is assigned a scalar design variable \( \rho_e \in [0,1] \), representing its material density. Here, \( \rho_e = 1 \) corresponds to solid material, and \( \rho_e = 0 \) denotes void. Intermediate values are allowed but typically penalized to promote binary solutions.

The standard optimization problem is formulated as:
\begin{equation}
\begin{aligned}
\min_{\rho} \quad & C(\rho) = \mathbf{u}^\top \mathbf{K}(\rho) \mathbf{u} \\
\text{s.t.} \quad & \mathbf{K}(\rho) \mathbf{u} = \mathbf{f}, \\
& \frac{1}{|\Omega|} \int_{\Omega} \rho(\mathbf{x}) \, d\mathbf{x} \leq V^*, \\
& 0 \leq \rho_e \leq 1 \quad \forall e,
\end{aligned}
\label{eq:compliance}
\end{equation}
where \( \mathbf{K}(\rho) \) is the global stiffness matrix assembled from element-wise contributions, \( \mathbf{u} \) is the displacement vector, \( \mathbf{f} \) is the force vector, and \( V^* \) is the maximum allowed volume fraction.

To relate material density to stiffness, the solid isotropic material with penalization (SIMP) method~\cite{sigmund2001} is widely adopted. In SIMP, the Young’s modulus of an element is interpolated as
\begin{equation}
E_e(\rho_e) = E_{\min} + \rho_e^p (E_0 - E_{\min}),
\label{eq:simp}
\end{equation}
where \( E_0 \) is the Young’s modulus of the fully solid material, \( E_{\min} \) is a small positive number to prevent singularities, and \( p \) is the penalization exponent, typically between 3 and 5. This interpolation discourages intermediate densities and drives the design toward black-and-white solutions.

The finite element method (FEM) provides the physical simulation engine for evaluating structural performance. For a given material distribution \( \rho = \{\rho_e\} \), the global stiffness matrix \( \mathbf{K}(\rho) \) is assembled using the penalized element stiffness matrices, and the equilibrium equation
\begin{equation}
\mathbf{K}(\rho) \mathbf{u} = \mathbf{f}
\end{equation}
is solved to obtain the displacement field \( \mathbf{u} \).
Once displacements are known, the structural compliance is evaluated as
\begin{equation}
C(\rho) = \sum_{e=1}^{N_e} \rho_e^p\, \mathbf{u}_e^\top \mathbf{K}_e^0 \mathbf{u}_e,
\label{eq:compliance_elementwise}
\end{equation}
where \( \mathbf{u}_e \) is the displacement vector associated with element \( e \), and \( \mathbf{K}_e^0 \) is the stiffness matrix of a fully solid element. This element-wise formulation highlights the contribution of each element to the overall structural performance.


To update the design in each iteration of topology optimization, gradient-based methods require information about how small changes in the design variables \( \rho_e \) affect the objective function, such as compliance. This is accomplished by computing the sensitivity of the objective with respect to each design variable. Sensitivities guide the optimizer in determining whether to increase or decrease material density in each element to reduce compliance while satisfying constraints.

In traditional FEM-based TO, sensitivities are computed using the adjoint method. The adjoint approach computes the full gradient vector at a cost equivalent to one additional FEM solve, regardless of the number of design variables. Therefore, each iteration typically involves two linear system solves, one for the forward displacement and one for the adjoint problem (which is often identical to the forward problem in compliance-based formulations).
However, this still represents a significant computational burden, especially in high-resolution 3D problems or multi-physics settings. Furthermore, deriving adjoint equations for new objectives, constraints, or coupled physical phenomena can be analytically challenging and prone to error.

In contrast, machine learning–based topology optimization methods, such as the approach proposed in this paper, eliminate the need for manually derived sensitivities by utilizing automatic differentiation (AD) within a neural network–driven design framework \cite{chandrasekhar2021tounn, shishir2024multi}. Since the material distribution is generated by a differentiable decoder and the objective function is incorporated into a continuous loss function, gradients with respect to the design parameters can be efficiently computed via backpropagation. This obviates the need to solve adjoint equations, significantly simplifying implementation and improving the flexibility of the method for extension to complex or multi-physics problems.

\section{Variational Topology Optimization Framework}

A central challenge in topology optimization is identifying a compact yet expressive representation of structural designs that can be effectively optimized using gradient-based methods. Traditional approaches typically rely on direct parameterization of the design domain via density variables assigned to each finite element. More recently, machine learning-based frameworks have introduced the use of latent vectors passed through neural decoders to generate spatially distributed material layouts.

Recent studies have explored the use of latent vector representations in topology optimization, particularly through autoencoders \cite{ma2023learning}, variational autoencoders (VAEs) \cite{nie2021topologygan, parekh2022variational}, and generative adversarial networks (GANs)~\cite{rawat2019deep, nie2021topologygan, kus2024gradientfree}. In these data-driven frameworks, a decoder network is trained to reconstruct or generate structural topologies from a low-dimensional latent space, often using thousands of precomputed samples generated by classical topology optimization solvers. Once trained, the latent space can be explored to interpolate between designs, accelerate prediction, or serve as a surrogate model. However, these approaches rely heavily on curated datasets and supervised training, which can be computationally expensive to generate and difficult to scale to new problem settings or boundary conditions.

In contrast, the method proposed in this work is entirely self-supervised and does not require any pre-existing dataset or offline training. The latent vector, whether sampled from a Gaussian distribution or produced by a variational quantum circuit, is optimized directly through physics-based objectives evaluated via finite element analysis. This formulation enables end-to-end topology optimization without relying on prior design data, and allows for adaptive exploration of the design space during optimization. As such, it offers a flexible and scalable alternative to existing data-driven generative design frameworks, while also introducing quantum encodings as a novel means of latent space parameterization.

An overview of the full pipeline architecture, followed by a detailed description of its key components are presented next. 
\begin{figure}[!tb]
    \centering
    \includegraphics[width=1\textwidth, trim=0 0 0 0, clip]{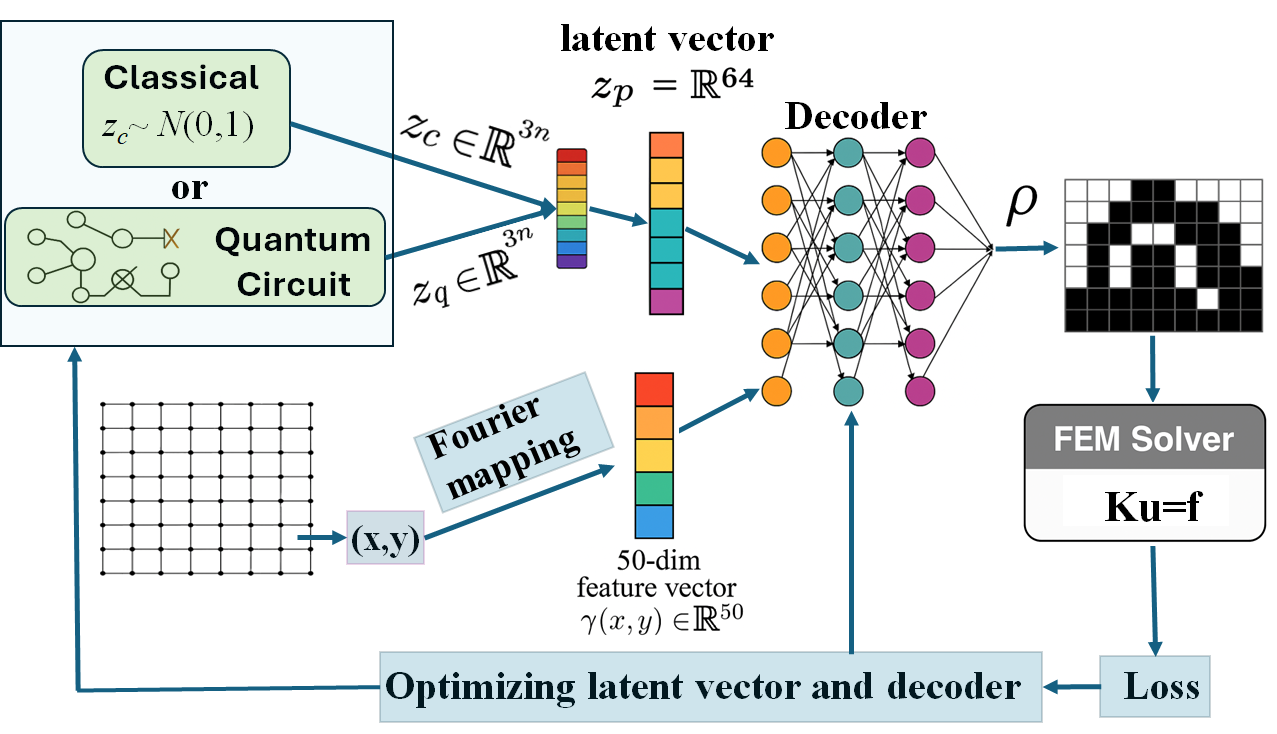}
\caption{Overview of the proposed variational topology optimization framework. A latent vector is generated using either a classical random initialization or a quantum circuit. It is projected to a higher-dimensional space and concatenated with Fourier-mapped spatial coordinates. The combined input is passed to a neural decoder that predicts the material distribution. The predicted density is evaluated using finite element analysis, and gradients are propagated through the entire model using automatic differentiation.}
    \label{fig:schem}
\end{figure}

\subsection{Pipeline Overview}

Figure~\ref{fig:schem} illustrates the overall architecture of the proposed variational topology optimization framework. The method formulates structural design as a variational learning problem in which a low-dimensional latent vector is decoded into a high-resolution material distribution. This pipeline is designed to support both quantum and classical latent encoding strategies, enabling direct comparison within a unified, self-supervised framework.

The process begins with the generation of a compact latent vector. In the classical setting, the vector is initialized by sampling from a standard Gaussian distribution. In the quantum setting, a variational quantum circuit (VQC) prepares an entangled multi-qubit state, from which expectation values of Pauli observables are measured to construct a structured, bounded latent vector. In both cases, the initial latent vector is low-dimensional (e.g., \( \mathbf{z}_q \in \mathbb{R}^9 \)) and is then projected to a higher-dimensional latent space (\( \mathbf{z}_p \in \mathbb{R}^{64} \)) using a trainable affine transformation.

To capture spatial variation, each point in the design domain is represented by its spatial coordinates, which are transformed using a Fourier feature mapping. This mapping enhances the input space with high-frequency components, enabling the decoder to represent complex geometry and fine-scale details. The projected latent vector is concatenated with the Fourier-encoded coordinates and passed to a fully connected neural decoder. The decoder maps this combined input to a scalar material density \( \rho(x, y) \in [0, 1] \), representing the local material distribution.

The predicted density field is used to assemble a finite element stiffness matrix and solve the linear elasticity problem. Structural performance is evaluated through compliance minimization under a volume constraint, with optional regularization terms for smoothness and binarization. All components of the pipeline are differentiable, allowing gradients to propagate through the decoder, projection layer, and quantum circuit (if used). The entire model is trained end-to-end using gradient-based optimization driven solely by physics-based loss functions, without any need for labeled datasets or supervised pretraining.

A central component of the proposed framework is the use of a compact latent vector to parameterize the structural design space. This latent representation serves as a global descriptor of the topology and is used by the decoder to reconstruct the full-resolution material distribution. By optimizing over this low-dimensional space, the method reduces the number of design variables and encourages structured solutions, while maintaining differentiability throughout the optimization process.

We explore two approaches for generating the latent vector: a quantum encoding based on a variational quantum circuit, and a classical encoding initialized from a standard normal distribution. Both methods produce a low-dimensional latent vector \( \mathbf{z}_q \in \mathbb{R}^k \) that is subsequently transformed via a trainable projection layer to a higher-dimensional latent code \( \mathbf{z}_p \in \mathbb{R}^{d_z} \), compatible with the decoder input.

The quantum encoder leverages the expressivity of entangled multi-qubit states and the bounded nature of expectation values to produce a compact, physically structured representation. In contrast, the classical latent vector offers a flexible but unstructured alternative that serves as a baseline for comparison. This design allows us to evaluate the relative advantages of quantum and classical encodings under a unified training framework.

The following subsections describe the details of both approaches, including circuit construction, parameter initialization, and training methodology.

\subsubsection{Quantum circuit for generating a latent vector}
We begin by describing the construction of the latent vector using a variational quantum circuit (VQC), which serves as a compact and expressive encoding mechanism within our topology optimization framework.
Quantum computing operates on quantum bits, or qubits, which unlike classical bits can exist in a superposition of states. A single qubit is represented by a normalized complex-valued vector in a two-dimensional Hilbert space. Its state can be expressed as
\[
|\psi\rangle = \alpha|0\rangle + \beta|1\rangle, \quad \text{where } \alpha, \beta \in \mathbb{C}, \quad |\alpha|^2 + |\beta|^2 = 1.
\]
This superposition property, when extended to multiple qubits, leads to an exponentially large state space of dimension \( 2^n \) for \( n \) qubits. Qubits may also become entangled, a uniquely quantum phenomenon in which the state of one qubit depends on the state of another, regardless of their spatial separation. Entanglement plays a crucial role in enabling quantum circuits to represent structured, non-separable latent information in a compact and trainable form.

\begin{figure}[!tb]
    \centering
    \includegraphics[width=\textwidth, trim=60 40 60 0, clip]{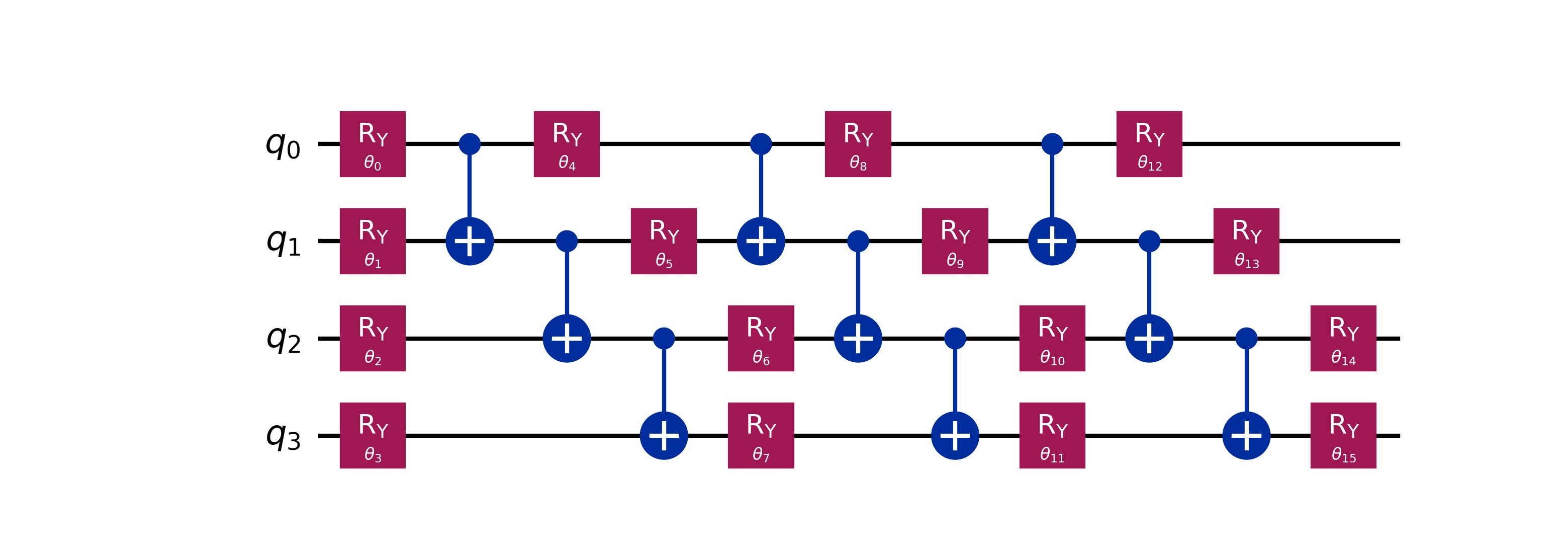}
    \caption{Quantum circuit used in the latent encoder. The circuit consists of layers of parameterized \( R_Y \) rotations and CNOT entanglement gates, repeated over multiple blocks. }
    \label{fig:quantum_latent_encoder}
\end{figure}

Our quantum latent encoder constructs a parameterized quantum circuit that acts on a register of \( n \) qubits, each initialized in the standard basis state \( |0\rangle \). The circuit, shown in Figure~\ref{fig:quantum_latent_encoder}, applies a sequence of unitary transformations to this initial state, governed by a set of trainable parameters \( \boldsymbol{\theta} \in \mathbb{R}^P \). The first layer of the circuit applies single-qubit rotations around the Y-axis of the Bloch sphere, denoted by the gate \( R_Y(\theta_i) \), which is defined by the matrix
\[
R_Y(\theta) = 
\begin{bmatrix}
\cos(\theta/2) & -\sin(\theta/2) \\
\sin(\theta/2) & \cos(\theta/2)
\end{bmatrix}.
\]
This gate transforms the qubit's state vector by a rotation parameterized by a real angle \( \theta \). 

Following the initial rotation layer, the circuit includes a block of entangling operations consisting of controlled-NOT (CNOT) gates applied between neighboring qubits. Each CNOT gate performs a conditional operation on a target qubit depending on the state of a control qubit and is responsible for creating multi-qubit entanglement, which significantly enhances the representational capacity of the quantum circuit. Another round of parameterized \( R_Y \) rotations follows, and this pattern—entanglement followed by rotation—is repeated for a user-defined number of repetitions. Each repetition adds an additional layer of expressivity to the circuit, allowing it to capture increasingly complex correlations across the qubits. As with deep neural networks, increasing the circuit depth via repetitions enhances the capacity of the quantum model while balancing training stability and parameter count.

In this work, the quantum circuit is restricted to \( R_Y \) rotations and entangling CNOT gates, excluding rotations about the X- or Z-axes. Although quantum states are generally complex-valued, this design ensures that all intermediate states remain real. Such a constraint is not restrictive; the combination of \( R_Y \) and CNOT gates is sufficient to generate any real-valued unitary transformation in the Hilbert space, thus retaining the expressivity needed for our application while avoiding unnecessary complexity in state evolution.
Maintaining real-valued quantum states offers several important advantages. From a computational standpoint, real amplitudes simplify simulation on classical hardware by reducing memory consumption and the cost of matrix operations. In our framework, where only real-valued expectation values are computed and passed to a classical decoder, phase information introduced by complex gates such as \( R_Z \) is mathematically valid but practically redundant. Including such phase dynamics not only increases the number of trainable parameters but also complicates the optimization landscape without delivering meaningful performance gains. In contrast, the real-valued architecture reduces parameter count and eliminates complex-valued operations, which facilitates more stable training and improves the efficiency of gradient-based updates, particularly when using the parameter-shift rule. Moreover, circuits composed solely of \( R_Y \) gates tend to exhibit smoother and more predictable convergence behavior compared to those involving general SU(2) rotations. Finally, single-axis rotations simplify circuit implementation and enhance compatibility with near-term quantum hardware, where minimizing gate depth and operational complexity is essential for mitigating noise and decoherence.
\subsection{Latent Vector Representations}
\label{sec:latent_vectors}

Formally, the full unitary transformation implemented by the variational quantum circuit is given by
\[
U(\boldsymbol{\theta}) = \left( \prod_{k=1}^{r} \left[ \bigotimes_{i=1}^{n} R_Y(\theta_{k,i}) \cdot \mathcal{E} \right] \right) \cdot \left( \bigotimes_{i=1}^{n} R_Y(\theta_{0,i}) \right),
\]
where \( \boldsymbol{\theta} \in \mathbb{R}^{n(r+1)} \) denotes the collection of all variational parameters, \( \mathcal{E} \) is the entangling layer composed of \( n-1 \) CNOT gates acting between adjacent qubits, and \( \bigotimes \) represents a tensor product across all qubits. The circuit prepares a quantum state
\[
|\psi(\boldsymbol{\theta})\rangle = U(\boldsymbol{\theta}) |0\rangle^{\otimes n},
\]
which encodes the learned quantum representation of the latent space.
Once the circuit has been executed, it prepares a quantum state \( |\psi(\boldsymbol{\theta})\rangle \) over the \( n \)-qubit register. To extract classical information from this state, we perform expectation value measurements of the standard Pauli observables \( Z \), \( X \), and \( Y \) for each qubit. These Hermitian operators are defined respectively as
\begin{equation}
Z = 
\begin{bmatrix}
1 & 0 \\
0 & -1
\end{bmatrix}, \quad
X = 
\begin{bmatrix}
0 & 1 \\
1 & 0
\end{bmatrix}, \quad
Y = 
\begin{bmatrix}
0 & -i \\
i & 0
\end{bmatrix}.    
\end{equation}
For each qubit \( i \) in an \( n \)-qubit system, we define a corresponding single-qubit observable \( P_i \) acting on the full Hilbert space as
\begin{equation}
    P_i = I^{\otimes i} \otimes P \otimes I^{\otimes(n - i - 1)},
\end{equation}
where \( P \in \{X, Y, Z\} \) is one of the Pauli matrices. Using this construction, we compute the expectation values
\[
x_i = \langle \psi(\boldsymbol{\theta}) | P_i | \psi(\boldsymbol{\theta}) \rangle \quad \text{with } P = X,
\]
\[
y_i = \langle \psi(\boldsymbol{\theta}) | P_i | \psi(\boldsymbol{\theta}) \rangle \quad \text{with } P = Y.
\]
\[
z_i = \langle \psi(\boldsymbol{\theta}) | P_i | \psi(\boldsymbol{\theta}) \rangle \quad \text{with } P = Z,
\]

These are then concatenated to form the quantum latent vector:
\begin{equation}
 \mathbf{z}_{\text{q}} =
\left[
\langle Z_0 \rangle, \dots, \langle Z_{n-1} \rangle,\;
\langle X_0 \rangle, \dots, \langle X_{n-1} \rangle,\;
\langle Y_0 \rangle, \dots, \langle Y_{n-1} \rangle
\right] \in \mathbb{R}^{3n},   
\end{equation}

where each component lies in the interval \([-1, 1]\). This vector compactly captures the geometry, coherence, and entanglement structure of the quantum state in a structured, differentiable, and numerically stable form suitable for downstream decoding.

\textbf{Latent Space Expansion via Learnable Projection:}
Because the decoder network that follows the quantum encoder typically requires a higher-dimensional latent input, the quantum latent vector is passed through a trainable linear projection layer. This projection also allows us to significantly reduce the number of physical qubits required in the circuit. Rather than increasing the qubit count to match the decoder's latent dimension, we maintain a compact quantum state (e.g., using 3–5 qubits) and expand its representation through the learnable projection. This not only reduces computational cost in simulation or on quantum hardware, but also promotes a more structured and constrained latent encoding that avoids overparameterization.

The projection layer transforms the quantum output vector \( \mathbf{z}_q \in \mathbb{R}^{3n} \), where \( n \) denotes the number of qubits and each qubit contributes three features (e.g., expectation values of Pauli \( X \), \( Y \), and \( Z \)), into a higher-dimensional latent representation \( \mathbf{z}_p \in \mathbb{R}^{d_z} \), with \( d_z > 3n \). This transformation is implemented via an affine map
\begin{equation}\label{eq:project}
\mathbf{z}_p = W \mathbf{z}_q + \mathbf{b},
\end{equation}
where \( W \in \mathbb{R}^{d_z \times 3n} \) and \( \mathbf{b} \in \mathbb{R}^{d_z} \) are learnable parameters. The projected vector \( \mathbf{z}_p \) captures a richer and more expressive encoding of the quantum information. The latent vector is then concatenated with Fourier-encoded spatial coordinates and passed as input to a coordinate-based neural decoder, which reconstructs the full-resolution material density field.


\textbf{Implementation and Training of the Quantum Latent Vector:}
The quantum encoder is implemented using PennyLane, a software framework for differentiable quantum programming, with PyTorch as the classical backend. This integration allows the quantum circuit to participate in PyTorch's automatic differentiation pipeline. Gradients of quantum expectation values with respect to circuit parameters are computed internally using the parameter-shift rule, which is a quantum-compatible generalization of the finite difference method. For a rotation gate parameter \( \theta_j \), and an observable \( \mathcal{O} \), the derivative is given by
\begin{equation}    
\frac{\partial \langle \mathcal{O} \rangle}{\partial \theta_j} = 
\frac{1}{2} \left(
\langle \mathcal{O} \rangle_{\theta_j + \frac{\pi}{2}} - 
\langle \mathcal{O} \rangle_{\theta_j - \frac{\pi}{2}}
\right).
\end{equation}
PennyLane performs these calculations efficiently and returns the results to PyTorch’s autograd engine, enabling the entire quantum-classical model to be trained end-to-end using gradient-based optimizers such as Adam or L-BFGS. 
Importantly, the entire latent-to-density pipeline is trained using only physics-based loss functions, typically structural compliance computed via finite element analysis. During each optimization iteration, the quantum parameters \( \boldsymbol{\theta} \), the projection matrix and other trainable components of the model are updated to minimize compliance under volume constraints. All gradients are computed automatically, and the entire process is fully self-supervised.



\subsubsection{Classical Latent Vector Generation}
To provide a baseline for comparison, we also consider a classical formulation in which the latent vector is initialized and optimized entirely within the classical domain. In this setting, the latent vector is a trainable tensor of fixed dimension \( d_z \), typically initialized by sampling from a standard Gaussian distribution, \( \mathcal{N}(0, 1) \). Unlike the quantum-generated latent space, which inherently produces structured and bounded features via entanglement and observable expectations, the classical latent vector is unstructured and unbounded—each component is a free parameter unconstrained by any physical model or encoding prior.

This lack of internal structure provides flexibility but also poses challenges. While an unbounded latent space may allow for greater expressivity, especially in high-dimensional regimes, it can lead to instability during training. Without architectural constraints or natural bounds, large or poorly scaled latent values may cause sharp transitions in the decoder output or impede convergence. Furthermore, because the classical latent vector lacks spatial correlation or physical grounding, it functions as a black-box parameterization, placing the burden of learning meaningful structure entirely on the decoder network.

To promote architectural consistency with the quantum approach and mitigate issues of overparameterization, we follow a similar pipeline by first generating a reduced-dimensional latent vector \( \mathbf{z}_c \in \mathbb{R}^{k} \), where \( k \ll d_z \), and then applying the same learnable affine projection defined in Eq.~\eqref{eq:project}, \( \mathbf{z}_p = W \mathbf{z}_c + \mathbf{b} \), to obtain the full decoder input \( \mathbf{z}_p \in \mathbb{R}^{d_z} \). This projection layer introduces an intermediate structured mapping that regularizes the optimization process and encourages smoother gradients by restricting the latent search space to a lower-dimensional, trainable manifold.

As in the quantum case, the projected latent vector is concatenated with Fourier-encoded spatial coordinates and passed to the coordinate-based decoder, which reconstructs the material density field. Optimization is performed end-to-end using physics-based loss functions such as compliance, evaluated via finite element analysis. All components, including the classical latent vector and decoder weights, are updated jointly through PyTorch's automatic differentiation engine.

\section{Fourier Feature Mapping for Coordinate-based Neural Fields}


The decoder used in our framework is a coordinate-based neural network that takes as input the spatial location and a global latent vector to predict the material density at each point. 

Coordinate-based neural networks, often referred to as neural fields or implicit neural representations, are increasingly used to model spatially varying functions such as material distributions, displacement fields, and scalar physical quantities~\cite{mildenhall2020nerf}. These models are typically defined as continuous mappings
\( f_\phi: \mathbb{R}^{d_x} \to \mathbb{R}\),
where \( \mathbf{x} \in \mathbb{R}^{d_x} \) denotes a spatial coordinate in a \( d_x \)-dimensional design domain (e.g., \( d_x = 2 \) for 2D problems), and the output \( f_\phi(\mathbf{x}) \) represents the predicted physical quantity at that location.
Standard multilayer perceptrons (MLPs), when applied directly to spatial coordinate regression, exhibit a phenomenon known as spectral bias—the tendency to learn low-frequency functions more easily than high-frequency ones~\cite{rahaman2019spectral}. As a consequence, such networks often struggle to represent sharp interfaces or fine-grained geometric features in the predicted field. This limitation can be mitigated by applying a Fourier feature mapping to the input coordinates prior to network evaluation~\cite{tancik2020fourier}. The inclusion of sinusoidal features has been shown to significantly enhance both convergence and accuracy in a variety of applications, including image reconstruction~\cite{tancik2020fourier}, neural radiance fields~\cite{mildenhall2020nerf}, and physics-informed neural operators~\cite{wang2021learning}. In the context of topology optimization, Fourier-mapped coordinates enable the decoder network to capture both global structure and localized detail without relying on spatial discretization or convolutional priors, thereby improving the expressiveness of the learned material distribution.

The Fourier feature mapping enriches the input space by encoding spatial coordinates \( \mathbf{x} \in \mathbb{R}^{d_x} \) using sinusoidal basis functions at multiple frequencies. Let \( \mathbf{B} \in \mathbb{R}^{m \times d_x} \) denote a matrix of frequency vectors, where each row \( \mathbf{b}_i \in \mathbb{R}^{d_x} \) defines a unique frequency and orientation, and \( m \) is the number of frequencies used. The transformed features are computed by applying sine and cosine functions to the inner products \( \mathbf{b}_i^\top \mathbf{x} \), yielding the Fourier embedding:
\begin{equation} \label{eq:fourier}
    \gamma(\mathbf{x}) = \left[ \sin\left(2\pi \mathbf{B} \mathbf{x} \right), \cos\left(2\pi \mathbf{B} \mathbf{x} \right) \right] \in \mathbb{R}^{2m}.
\end{equation}
This formulation introduces a set of oscillatory basis functions that probe the input domain at varying resolutions and orientations. In our setting, the Fourier-mapped coordinates \( \gamma(\mathbf{x}) \) are concatenated with the latent design vector \( \mathbf{z}_p \in \mathbb{R}^{d_z} \) and passed into a coordinate-based neural decoder. 


The frequency matrix \( \mathbf{B} \) plays a central role in determining the expressive capacity of the mapping. One common strategy is to sample each row of \( \mathbf{B} \) from an isotropic Gaussian distribution:
\begin{equation}\label{eq:gaussian_sampling}
    \mathbf{b}_i \sim \mathcal{N}(\mathbf{0}, \sigma^2 \mathbf{I}),
\end{equation}
where the variance \( \sigma^2 \) controls the spectral bandwidth of the encoding. Higher values of \( \sigma \) introduce higher frequency components and allow the network to resolve finer spatial features. 
Beyond random Gaussian sampling, alternative frequency selection strategies have been proposed. Logarithmically spaced frequencies, as employed in neural radiance fields (NeRF), systematically cover a range of scales, facilitating the capture of both coarse and fine details~\cite{mildenhall2020nerf}. Another approach involves making the frequency matrix \( \mathbf{B} \) learnable, allowing the model to adaptively select frequencies that best represent the data during training~\cite{li2021learnable}. This method introduces additional parameters but can enhance performance in tasks requiring precise frequency localization. Additionally, axis-aligned frequencies, where each \( \mathbf{b}_i \) is aligned with a single coordinate axis, can simplify the model and reduce computational complexity, albeit at the potential cost of expressiveness. 

In our framework, each spatial coordinate \( \mathbf{x} \in \mathbb{R}^{d_x} \) corresponds to the center of a finite element in the design domain. We apply a band-limited Fourier mapping to \( \mathbf{x} \), projecting it into a higher-dimensional periodic basis composed of sine and cosine functions at multiple frequencies. The frequency magnitudes are sampled uniformly from a bounded range and assigned random signs to ensure directional diversity. The resulting frequency-encoded features are concatenated with the latent design vector and passed to the neural decoder. This input representation endows the model with an inductive bias toward capturing fine-scale, high-frequency variations in the material distribution. For efficiency and stability, the frequency matrix is sampled once at initialization and held fixed throughout training, forming a static but expressive basis for spatial encoding without introducing additional learnable parameters.

\subsection{Decoder}\label{decoder}
The mapping from latent and spatial inputs to material density is performed by a neural decoder, as illustrated in~\fref{fig:schem}. The decoder function \( f_\phi \), implemented as a fully connected multilayer perceptron (MLP) with parameters \( \phi \), maps the concatenation of the projected latent vector \( \mathbf{z}_p \in \mathbb{R}^{d_z} \) and the Fourier-encoded spatial coordinate \( \gamma(\mathbf{x}) \in \mathbb{R}^{d_\gamma} \) to a scalar density value:
\begin{equation}
    \rho(\mathbf{x}) = f_\phi\left( \mathbf{z}_p, \gamma(\mathbf{x}) \right), \quad \rho(\mathbf{x}) \in [0,1].
\end{equation}

The latent vector \( \mathbf{z}_p \) encodes global design information and is shared across all spatial locations, while the Fourier-mapped coordinate \( \gamma(\mathbf{x}) \) encodes local spatial features. To evaluate the density field over the domain, \( \mathbf{z}_p \) is broadcasted and concatenated with \( \gamma(\mathbf{x}) \) at each location, forming a composite input \( \mathbf{u} = [\mathbf{z}_p, \gamma(\mathbf{x})] \in \mathbb{R}^{d_z + d_\gamma} \).

The network architecture consists of four fully connected layers. The first layer maps the input \( \mathbf{u} \) to a hidden representation of size 64, followed by a ReLU activation. This is followed by an expansion to 128 neurons and a subsequent contraction back to 64, each with ReLU activations. The final layer reduces the output to a scalar, which is passed through a sigmoid activation to ensure \( \rho(\mathbf{x}) \in (0,1) \), enabling continuous and differentiable predictions suitable for gradient-based optimization.

While the model is relatively lightweight compared to deep convolutional networks, its coordinate-based design provides high spatial resolution and flexibility. By conditioning on both global and local information, the decoder can represent complex topologies without relying on mesh-based parameterization. The decoder is trained jointly with the latent encoding and physics loss, with its parameters \( \phi \) updated via backpropagation from a compliance-driven loss function augmented with volume and regularization penalties.
\subsection{Regularization Strategies and Density Sharpening}

In classical topology optimization, particularly within the SIMP framework, regularization techniques such as density filtering \cite{bendsoe2003topology} and Heaviside projection \cite{guest2004achieving} are commonly employed to mitigate numerical instabilities and encourage binary material layouts. Density filtering smooths the density field via local averaging (e.g., using a circular or Gaussian kernel), helping to eliminate checkerboarding and mesh dependency~\cite{sigmund2007morphology}. However, its non-local nature introduces additional computational overhead and complicates gradient flow when used within a neural network framework.
Heaviside projection aims to push intermediate densities toward binary values by applying a sharp, parameterized function to the filtered output~\cite{guest2004achieving}. While effective, this approach can introduce non-smooth gradients and often requires continuation schemes to gradually increase the projection slope over time, making training more complex and expensive.

In our framework, we experimented with both filtering and Heaviside projection applied externally to the decoder output. These methods improved binary convergence but significantly slowed training and increased computational cost, especially in high-resolution cases, due to the need for non-local operations and custom gradient propagation.

To address these limitations while still promoting binarization, we adopt a differentiable approximation of the Heaviside function directly within the decoder. Specifically, we apply a sharp sigmoid activation to the output density
\begin{equation}\label{proj1}
    \rho_\beta(\mathbf{x}) = \frac{1}{1 + \exp\left( -\beta \left( \rho(\mathbf{x}) - 0.5 \right) \right)},
\end{equation}
where \( \beta > 0 \) controls the steepness of the transition. To progressively enforce near-binary values during training, we apply a continuation scheme in which \( \beta \) increases from 1.0 to a maximum value of 10.0. 
This ensures a gradual sharpening of the density field while preserving differentiability and numerical stability throughout the optimization.

In addition to output sharpening, we incorporate explicit regularization terms in the loss function to guide the solution toward feasible and physically meaningful topologies. The loss function, including the optional smoothness-promoting penalties, is described in the following subsection.

\subsection{Loss function}\label{loss}
The projected density field \( \rho_\beta \) is used to assemble the global stiffness matrix \( \mathbf{K}(\rho_\beta) \), and the structural displacements \( \mathbf{u} \in \mathbb{R}^{n_u} \) are computed by solving the linear system
\begin{equation}
    \mathbf{K}(\rho_\beta) \, \mathbf{u} = \mathbf{f},
\end{equation}
where \( \mathbf{f} \) denotes the global load vector. The structural compliance is evaluated as
\begin{equation}
    C(\rho_\beta) = \mathbf{u}^\top \mathbf{K}(\rho_\beta) \mathbf{u}.
\end{equation}

To enforce a prescribed material usage, we introduce a volume constraint that penalizes deviations from a target volume fraction \( V^* \in (0,1) \). Let \( \Omega \subset \mathbb{R}^{d_x} \) denote the design domain, and let \( |\Omega| \) represent its Lebesgue measure (i.e., its total area in 2D or volume in 3D). The average material density over the domain is given by
\begin{equation}
    \bar{\rho} = \frac{1}{|\Omega|} \int_\Omega \rho_\beta(\mathbf{x}) \, d\mathbf{x},
\end{equation}
where \( \rho(\mathbf{x}) \in [0,1] \) denotes the predicted density at location \( \mathbf{x} \). The volume constraint is incorporated into the loss function as a quadratic penalty
\begin{equation}
    \mathcal{L}_{\text{vol}} = \lambda_{\text{vf}} \left( \bar{\rho} - V^* \right)^2,
\end{equation}
where \( \lambda_{\text{vf}} \) is a weighting parameter that increases during training, allowing the optimizer to first explore the design space freely and then progressively enforce the material constraint more strictly.
To further regularize the density field and improve numerical stability, we introduce several auxiliary penalties.

\subsubsection{Binarization Penalty for Eliminating Intermediate Densities}
While volume constraints help control the overall material usage, they do not directly encourage the material distribution to converge to binary (0–1) values. To promote crisp, interpretable designs, we introduce an additional regularization term that penalizes intermediate densities and favors solutions where \( \rho(\mathbf{x}) \in \{0,1\} \). This is achieved through a binarization penalty defined as:
\begin{equation}
    \mathcal{L}_{\text{bin}} = \lambda_{\text{bin}} \int_\Omega \rho(\mathbf{x}) \left( 1 - \rho(\mathbf{x}) \right) \, d\mathbf{x},
\end{equation}
where \( \lambda_{\text{bin}} > 0 \) is a tunable penalty weight. The integrand \( \rho(\mathbf{x})(1 - \rho(\mathbf{x})) \) attains its maximum at \( \rho = 0.5 \) and vanishes at the binary endpoints \( \rho = 0 \) and \( \rho = 1 \), thereby discouraging intermediate values across the domain.

In practice, the binarization loss is computed over a discretized mesh of \( N \) finite elements, where \( \rho_i \in [0,1] \) denotes the predicted density at the center of element \( i \). Assuming uniform element volume, the discrete form of the loss becomes:
\begin{equation}
    \mathcal{L}_{\text{bin}} \approx \lambda_{\text{bin}} \cdot \frac{|\Omega|}{N} \sum_{i=1}^{N} \rho_i \left( 1 - \rho_i \right).
\end{equation}

This term serves as a soft constraint that complements the sharp sigmoid projection applied at the decoder output. While the sigmoid activation enhances local sharpness in the predicted density field, the binarization loss promotes global consistency by penalizing persistent intermediate (grey) regions throughout the design domain. In practice, this regularization term is introduced progressively during training; it remains inactive in the early stages to allow broader exploration of the design space, and its influence is gradually increased in later epochs to encourage convergence toward a discrete, binary topology. 

\subsubsection{Total Variation Regularization for Spatial Smoothness}
To promote spatial coherence and suppress spurious oscillations in the predicted density field, we incorporate a total variation (TV) penalty into the loss function. This regularization term encourages piecewise-smooth solutions by penalizing the magnitude of local gradients and enforcing edge regularity across the design domain. The continuous form of the TV functional is given by
\begin{equation}
    \mathcal{L}_{\text{TV}} = \lambda_{\text{TV}} \int_\Omega \left( \left| \nabla_x \rho(\mathbf{x}) \right| + \left| \nabla_y \rho(\mathbf{x}) \right| \right) \, d\mathbf{x},
\end{equation}
where \( \lambda_{\text{TV}} > 0 \) is a tunable penalty weight and \( \nabla_x \rho \), \( \nabla_y \rho \) denote the partial derivatives of \( \rho \) in each spatial direction. This anisotropic form approximates the full isotropic TV norm, defined as
\begin{equation}
    \text{TV}(\rho) = \int_\Omega \left| \nabla \rho(\mathbf{x}) \right| \, d\mathbf{x},
\end{equation}
which measures the total magnitude of spatial variations in the density field.

On a regular 2D Cartesian grid with element spacings \( \Delta x \) and \( \Delta y \), the TV norm can be discretized using forward finite differences. Letting \( \rho_{i,j} \) denote the density value at grid index \( (i,j) \), the discrete form becomes
\begin{equation}
    \text{TV}(\rho) \approx \sum_{i=1}^{n_x-1} \sum_{j=1}^{n_y} \Delta y \, \left| \frac{\rho_{i+1,j} - \rho_{i,j}}{\Delta x} \right|
    + \sum_{i=1}^{n_x} \sum_{j=1}^{n_y-1} \Delta x \, \left| \frac{\rho_{i,j+1} - \rho_{i,j}}{\Delta y} \right|.
\end{equation}
This formulation provides an efficient and consistent estimate of the total variation for structured domains, and is straightforward to implement with modern autodifferentiation frameworks.

For unstructured meshes, such as triangulated surfaces or tetrahedral grids, finite difference approximations are not directly applicable. Instead, the TV regularization must be computed using a variational formulation based on mesh connectivity. A commonly used expression in this setting is a weighted edge-sum representation
\begin{equation}
    \text{TV}(\rho) = \sum_{(i,j) \in \mathcal{E}} w_{ij} \left| \rho_i - \rho_j \right|,
\end{equation}
where \( \mathcal{E} \) is the set of edges connecting adjacent nodes or elements, and \( w_{ij} > 0 \) is a geometric weight, typically based on edge length, shared face area, or cotangent Laplacian coefficients. This discrete formulation generalizes the concept of total variation to arbitrary mesh topologies while preserving key properties such as convexity and sparsity of the regularization term.

The TV penalty helps to balance sharp feature preservation with smooth transitions in the material layout, making it particularly effective when combined with other constraints such as volume control and binarization.

\subsection{Sobolev $H^1$ Regularization for Global Smoothness}
To promote global smoothness in the material distribution and suppress high-frequency oscillations, we include a Sobolev-type \( H^1 \) seminorm penalty in the loss function. This regularization term penalizes the squared gradient magnitude of the density field and favors solutions with continuous spatial variations. The continuous form of the penalty is given by
\begin{equation}
    \mathcal{L}_{H^1} = \lambda_{H^1} \int_\Omega \left( \left( \nabla_x \rho(\mathbf{x}) \right)^2 + \left( \nabla_y \rho(\mathbf{x}) \right)^2 \right) \, d\mathbf{x},
\end{equation}
where \( \lambda_{H^1} > 0 \) is a regularization parameter, and \( \nabla_x \rho \), \( \nabla_y \rho \) denote partial derivatives of the predicted density \( \rho \) along each coordinate direction.

On structured grids, the \( H^1 \) seminorm can be discretized using finite differences. Letting \( \rho_{i,j} \) represent the density at grid point \( (i,j) \), and assuming uniform spacings \( \Delta x \) and \( \Delta y \), the discrete approximation becomes
\begin{equation}
\label{eq:h1_structured}
    \mathcal{L}_{H^1} \approx \sum_{i=1}^{n_x-1} \sum_{j=1}^{n_y} \Delta y \, \left( \frac{\rho_{i+1,j} - \rho_{i,j}}{\Delta x} \right)^2
    + \sum_{i=1}^{n_x} \sum_{j=1}^{n_y-1} \Delta x \, \left( \frac{\rho_{i,j+1} - \rho_{i,j}}{\Delta y} \right)^2.
\end{equation}
This forward-difference formulation estimates the squared norm of the spatial gradient over each grid cell, providing a simple and efficient implementation for structured domains.

For unstructured meshes, a variational formulation based on finite element basis functions can be used. Let \( \rho_h(\mathbf{x}) = \sum_i \rho_i \phi_i(\mathbf{x}) \) denote the finite element approximation of the density field, where \( \phi_i(\mathbf{x}) \) are piecewise-linear nodal basis functions and \( \rho_i \) are nodal values. The continuous \( H^1 \) seminorm then becomes
\begin{equation}
    \|\rho_h\|^2_{H^1(\Omega)} = \int_\Omega \left| \nabla \rho_h(\mathbf{x}) \right|^2 \, d\mathbf{x} = \sum_{i,j} \rho_i \rho_j \int_\Omega \nabla \phi_i(\mathbf{x}) \cdot \nabla \phi_j(\mathbf{x}) \, d\mathbf{x}.
\end{equation}
This expression can be compactly written in matrix form as
\begin{equation}
    \mathcal{L}_{H^1} = \boldsymbol{\rho}^\top \mathbf{L} \boldsymbol{\rho},
\end{equation}
where \( \boldsymbol{\rho} \) is the vector of nodal densities and \( \mathbf{L} \) is the global stiffness matrix assembled from the diffusion terms of each element. This formulation is fully mesh-aware and aligns the regularization with the geometry and connectivity of the computational domain, ensuring consistent gradient penalization even on irregular or adaptive meshes.

The \( H^1 \) regularization is particularly effective in smoothing out fine-scale fluctuations in the density field while preserving the global structure of the design. When combined with total variation and binarization terms, it contributes to the generation of physically meaningful and numerically stable topologies.

The total loss function minimized during training is therefore
\begin{equation}
    \mathcal{L}(\rho) = \frac{C(\rho)}{C_0} + \mathcal{L}_{\text{vol}} + \mathcal{L}_{\text{bin}} + \mathcal{L}_{\text{TV}} + \mathcal{L}_{H^1},
\end{equation}
where \( C_0 \) denotes the compliance at the initial iteration, used for normalization. The optimization proceeds via backpropagation through all differentiable components of the pipeline, including the latent parameters, projection layer, neural decoder, and quantum circuit parameters (if present). All parameters are jointly updated using the Adam optimizer, with individual learning rates assigned to each parameter group.

\subsection{Gradient Flow for Quantum and Classical Latent Optimization}

Gradient-based optimization in our framework is enabled by the fully differentiable structure of the variational decoding pipeline. For both the quantum and classical cases, gradients of the loss function \( \mathcal{L} \), which depends on the sharpened density field \( \rho_\beta \), are propagated through the decoder, projection layer, and latent vector using the chain rule. Specifically, the dependency on the quantum circuit parameters \( \boldsymbol{\theta} \) is given by
\[
\frac{\partial \mathcal{L}}{\partial \boldsymbol{\theta}} =
\frac{\partial \mathcal{L}}{\partial \rho_\beta}
\cdot
\frac{\partial \rho_\beta}{\partial \rho}
\cdot
\frac{\partial \rho}{\partial \mathbf{z}_p}
\cdot
\frac{\partial \mathbf{z}_p}{\partial \mathbf{z}_q}
\cdot
\frac{\partial \mathbf{z}_q}{\partial \boldsymbol{\theta}},
\]
where \( \rho \) is the raw decoder output and \( \rho_\beta \) is the projected density obtained via the sharp sigmoid function. The term \( \partial \rho_\beta / \partial \rho \) reflects the slope of the sigmoid and ensures that binarization remains differentiable. The Jacobian \( \partial \mathbf{z}_p / \partial \mathbf{z}_q \) corresponds to the projection matrix \( W \), and the final term \( \partial \mathbf{z}_q / \partial \boldsymbol{\theta} \) is computed using the parameter-shift rule implemented in PennyLane.

For the classical latent vector \( \mathbf{z}_c \in \mathbb{R}^{n_c} \), the chain rule simplifies to
\[
\frac{\partial \mathcal{L}}{\partial \mathbf{z}_c} =
\frac{\partial \mathcal{L}}{\partial \rho_\beta}
\cdot
\frac{\partial \rho_\beta}{\partial \rho}
\cdot
\frac{\partial \rho}{\partial \mathbf{z}_p}
\cdot
\frac{\partial \mathbf{z}_p}{\partial \mathbf{z}_c},
\]
where the projection is defined as \( \mathbf{z}_p = W \mathbf{z}_c + \mathbf{b} \), and \( \partial \mathbf{z}_p / \partial \mathbf{z}_c = W \). All other gradients are handled automatically by PyTorch’s autograd engine. Unlike the quantum case, no specialized quantum gradient computation is needed, making the classical optimization path computationally efficient and implementation-friendly. While the classical latent space lacks the structured entanglement and boundedness of its quantum counterpart, it offers rapid convergence and reduced overhead, serving as a strong baseline for comparison within the hybrid optimization framework.

 \subsection{Gradient Flow for Quantum Latent Optimization}
The gradient of the loss function with respect to the quantum circuit parameters \( \boldsymbol{\theta} \) is computed using the chain rule, which propagates derivatives through each stage of the model. Specifically, since the loss depends on the sharpened density field \( \rho_\beta \), the full dependency of \( \mathcal{L} \) on \( \boldsymbol{\theta} \) is given by
\[
\frac{\partial \mathcal{L}}{\partial \boldsymbol{\theta}} =
\frac{\partial \mathcal{L}}{\partial \rho_\beta}
\cdot
\frac{\partial \rho_\beta}{\partial \rho}
\cdot
\frac{\partial \rho}{\partial \mathbf{z}_p}
\cdot
\frac{\partial \mathbf{z}_p}{\partial \mathbf{z}_q}
\cdot
\frac{\partial \mathbf{z}_q}{\partial \boldsymbol{\theta}},
\]
where \( \rho \) is the raw decoder output, and \( \rho_\beta \) is the projected density obtained by applying the sharp sigmoid function. The term \( \frac{\partial \rho_\beta}{\partial \rho} \) captures the local slope of the sigmoid and ensures that the sharpening transformation remains differentiable. The Jacobian \( \frac{\partial \mathbf{z}_p}{\partial \mathbf{z}_q} \) corresponds to the projection matrix \( W \), and the final term \( \frac{\partial \mathbf{z}_q}{\partial \boldsymbol{\theta}} \) is computed using the parameter-shift rule as implemented in PennyLane. This fully differentiable structure enables seamless end-to-end gradient-based optimization across quantum and classical components.

\subsection{Gradient Flow for Classical Latent Optimization}
To optimize the classical latent vector \( \mathbf{z}_c \in \mathbb{R}^{n_c} \), we rely on end-to-end automatic differentiation through the variational decoding pipeline. Since the classical latent vector is a direct input to the learnable projection layer, the dependency of the loss function \( \mathcal{L} \) on \( \mathbf{z}_c \) is governed by a straightforward application of the chain rule:
\begin{equation}
\frac{\partial \mathcal{L}}{\partial \mathbf{z}_c} =
\frac{\partial \mathcal{L}}{\partial \rho_\beta}
\cdot
\frac{\partial \rho_\beta}{\partial \rho}
\cdot
\frac{\partial \rho}{\partial \mathbf{z}_p}
\cdot
\frac{\partial \mathbf{z}_p}{\partial \mathbf{z}_c},
\end{equation}
where \( \rho \) denotes the raw output of the decoder and \( \rho_\beta \) is the sharpened (projected) density obtained via the sigmoid-based activation. The projection step is given by \( \mathbf{z}_p = W \mathbf{z}_c + \mathbf{b} \), making the Jacobian \( \partial \mathbf{z}_p / \partial \mathbf{z}_c \) equal to the projection matrix \( W \). The remaining derivatives are handled automatically via PyTorch’s autograd engine.
Unlike the quantum case, classical latent optimization does not require the use of the parameter-shift rule or any circuit-specific gradient computations, as the entire path from latent vector to loss is purely classical and differentiable.

This differentiable formulation supports resolution-independent decoding, accommodates both classical and quantum encodings, and integrates advanced regularization techniques to yield high-quality, binary, and physically feasible topologies.
\section{Numerical results}
In this section, we present a series of numerical experiments to evaluate the effectiveness of the proposed variational topology optimization framework. Our goal is to demonstrate the capability of the decoder-based formulation to generate high-resolution, physically feasible, and nearly binary topologies under various configurations. We investigate the impact of different latent encodings (classical vs. quantum), compare the performance of regularization strategies, and examine the effects of enforcing sharp transitions through projection and filtering.

All simulations are conducted on two-dimensional structural domains under static loading conditions, with the material behavior governed by linear elasticity. 
The design domain is discretized using a structured grid of $60\times30$
 bilinear quadrilateral (Q4) finite elements. 
 For numerical evaluation, the material parameters are nondimensionalized with \( E = 1 \) and \( \nu = 0.3 \), and the target volume fraction is set to \( v_0 = 0.4 \), limiting the amount of material allowed in the design domain.
 The loss function includes compliance, volume, and smoothness terms, and all parameters are optimized jointly using gradient-based methods. To adapt the optimization behavior throughout the training process, each penalty coefficient is scheduled to evolve over time. 
The volume penalty coefficient \( \lambda_{\mathrm{vf}} \) increases linearly from a minimum value of 1.0 to a maximum value of \( \lambda_{\mathrm{vf}}^{\mathrm{max}} = 50.0 \) over the first 25\% of training epochs, thereby gradually enforcing the volume constraint. 
The binary penalty term, which promotes discreteness by pushing intermediate densities toward 0 or 1, is activated only during the final 25\% of training. 
During this phase, the coefficient \( \lambda_{\mathrm{bin}} \) increases linearly from 0 to a maximum value of \( \lambda_{\mathrm{bin}}^{\mathrm{max}} = 5.0 \). 
The total variation regularization weight \( \lambda_{\mathrm{TV}} \), introduced to suppress numerical instabilities such as checkerboarding and to encourage local smoothness, grows linearly from 0 to a maximum of 0.015. 
In contrast, the Sobolev-type \( H^1 \) regularization coefficient \( \lambda_{H^1} \), which enforces global smoothness, decreases linearly from an initial value of 0.05 to a minimum value of 0.005 over the course of training, allowing the design to become increasingly detailed in later iterations. 

To evaluate the effect of latent vector generation on topology optimization outcomes, we compare quantum and classical encoding strategies. For the quantum encoding, we employ parameterized quantum circuits with either 3 or 5 qubits, each repeated through 5 layers of entangling operations. The latent representation is formed by measuring the expectation values of Pauli operators, resulting in a $3n$-dimensional vector (where $n$ is the number of qubits). This quantum latent vector is then upscaled to a 64-dimensional code using a trainable linear projection.
For a fair comparison, the classical encoding also produces a $3n$-dimensional latent vector by sampling from a standard Gaussian distribution, which is similarly mapped to 64 dimensions through a learnable affine transformation. 
 In both cases, the resulting 64-dimensional latent vector are passed through an identical decoder architecture consisting of fully connected layers with ReLU activations, which generates a continuous material density field over the design domain.
To account for stochastic variability, the topology optimization is independently repeated over 10 runs for each encoding scheme.

\subsubsection*{Design diversity metric}

To quantify the variability among topologies generated from independent runs of the optimization process, we introduce an average pairwise distance metric based on the discrete \( \ell_2 \) norm. This metric captures how structurally different the resulting topologies are within a given encoding scheme (classical or quantum), providing a measure of design diversity.

Let \( \rho^{(1)}, \rho^{(2)}, \dots, \rho^{(n)} \) denote the set of optimized topologies obtained from \( n \) independent runs of the optimization process. Each \( \rho^{(i)} \in \mathbb{R}^m \) represents the flattened (vectorized) form of the \( i \)-th topology, where \( m \) is the total number of pixels in the discretized design domain. To quantify the variability among these solutions, we define the average pairwise distance as
\begin{equation}
D_{\text{avg}} = \frac{2}{n(n - 1)} \sum_{1 \leq i < j \leq n} \left\| \rho^{(i)} - \rho^{(j)} \right\|_2,
\label{eq:diversity}
\end{equation}
where \( \| \cdot \|_2 \) denotes the Euclidean (i.e., \( \ell_2 \)) norm. This scalar metric reflects the average structural dissimilarity between all unique pairs of topologies within the set. A higher value of \( D_{\text{avg}} \) indicates greater diversity in the generated designs, while a lower value suggests convergence toward similar solutions.


\begin{figure}[!tb]
    \centering

    \begin{subfigure}[t]{0.45\textwidth}
        \centering
        \includegraphics[width=\textwidth]{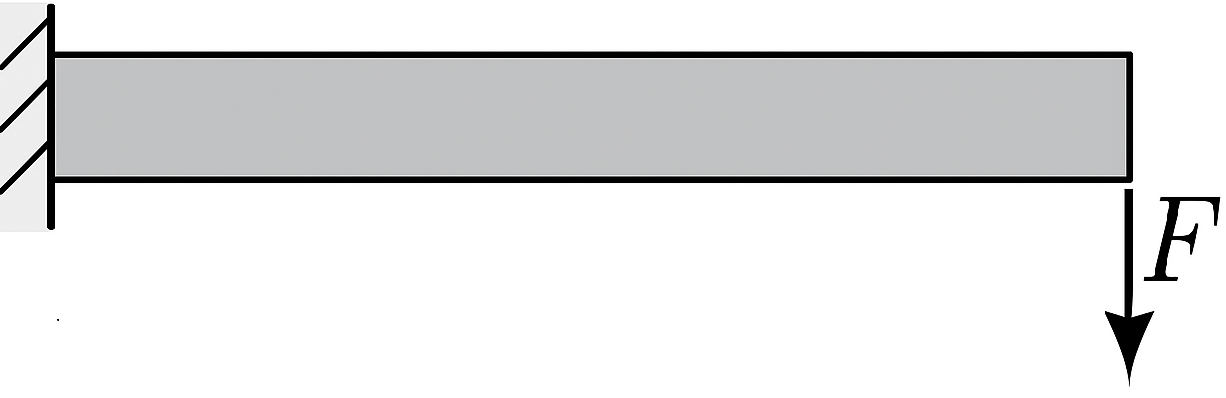}
        \caption{Cantilever beam setup}
    \end{subfigure}

    \vspace{0.5cm}

    \par\medskip
    \begin{subfigure}[t]{0.3\textwidth}
        \centering
        \includegraphics[width=\textwidth]{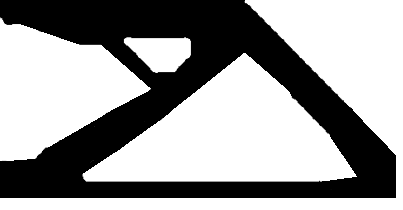}
        \caption{Compliance = 89.33}
    \end{subfigure}
    \hfill
    \begin{subfigure}[t]{0.3\textwidth}
        \centering
        \includegraphics[width=\textwidth]{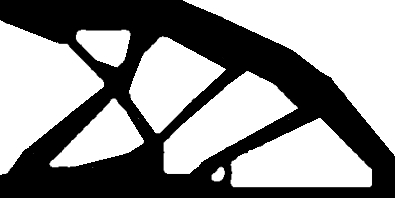}
        \caption{Compliance = 85.87}
    \end{subfigure}
    \hfill
    \begin{subfigure}[t]{0.3\textwidth}
        \centering
        \includegraphics[width=\textwidth]{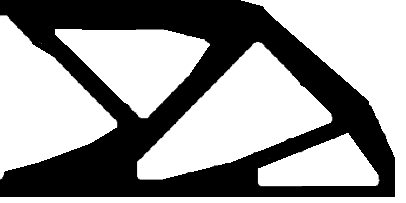}
        \caption{Compliance = 87.39}
    \end{subfigure}

    \par\medskip
    \begin{subfigure}[t]{0.3\textwidth}
        \centering
        \includegraphics[width=\textwidth]{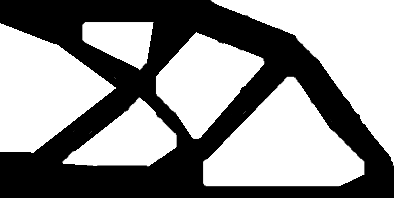}
        \caption{Compliance = 86.05}
    \end{subfigure}
    \hfill
    \begin{subfigure}[t]{0.3\textwidth}
        \centering
        \includegraphics[width=\textwidth]{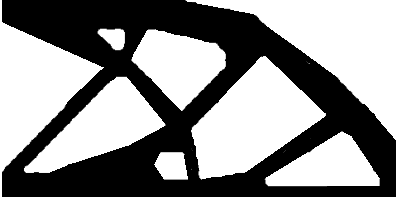}
        \caption{Compliance = 87.10}
    \end{subfigure}
    \hfill
    \begin{subfigure}[t]{0.3\textwidth}
        \centering
        \includegraphics[width=\textwidth]{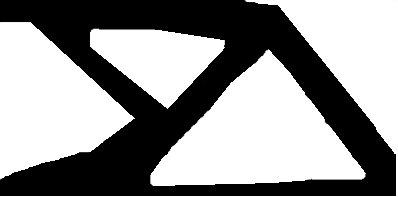}
        \caption{Compliance = 88.98}
    \end{subfigure}

    \caption{Comparison of topologies generated for a tip-loaded cantilever beam using classical and quantum latent encodings. The top row shows the geometry and boundary/loading conditions. Compliance values are listed below each topology.}
    \label{fig:topology_comparison}
\end{figure}
\subsection{Tip load cantilever beam}

\begin{figure}[!tb]
    \centering
    \begin{subfigure}[t]{0.49\textwidth}
        \centering
        \includegraphics[height=4.5cm]{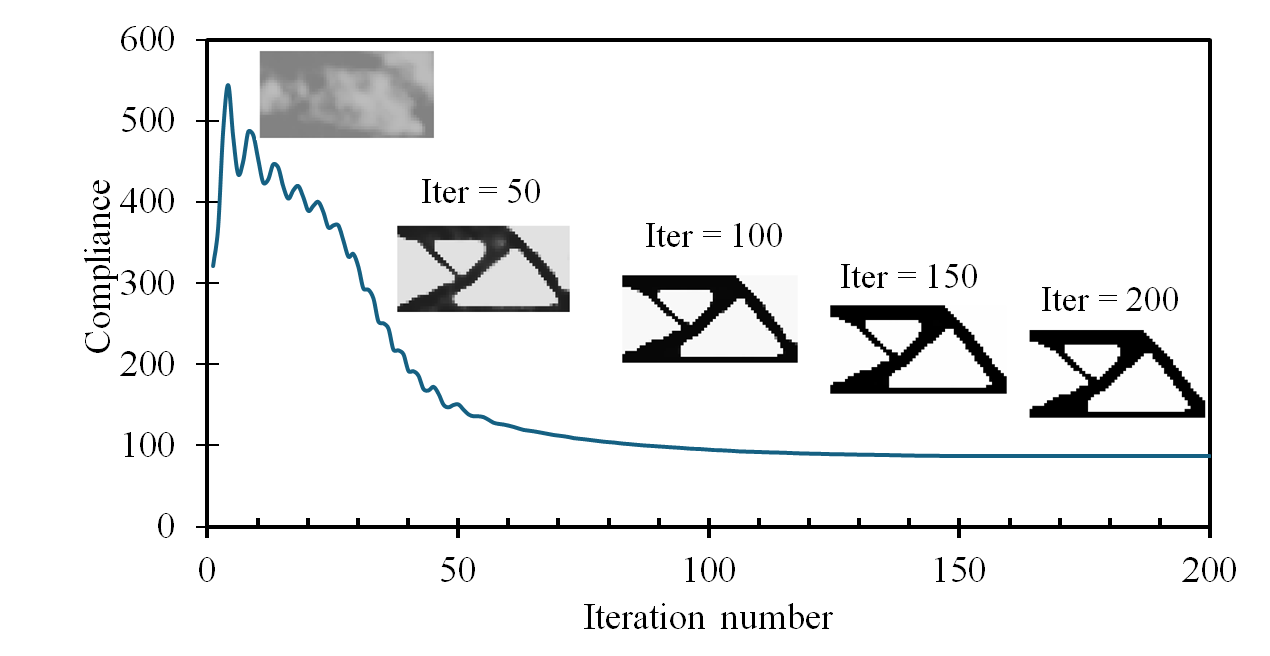}
        \caption*{\small Quantum latent vector, Compliance = 85.72}
        \label{fig:compliance-noqubit5}
    \end{subfigure}
    \hfill
    \begin{subfigure}[t]{0.49\textwidth}
        \centering
        \includegraphics[height=4.5cm]{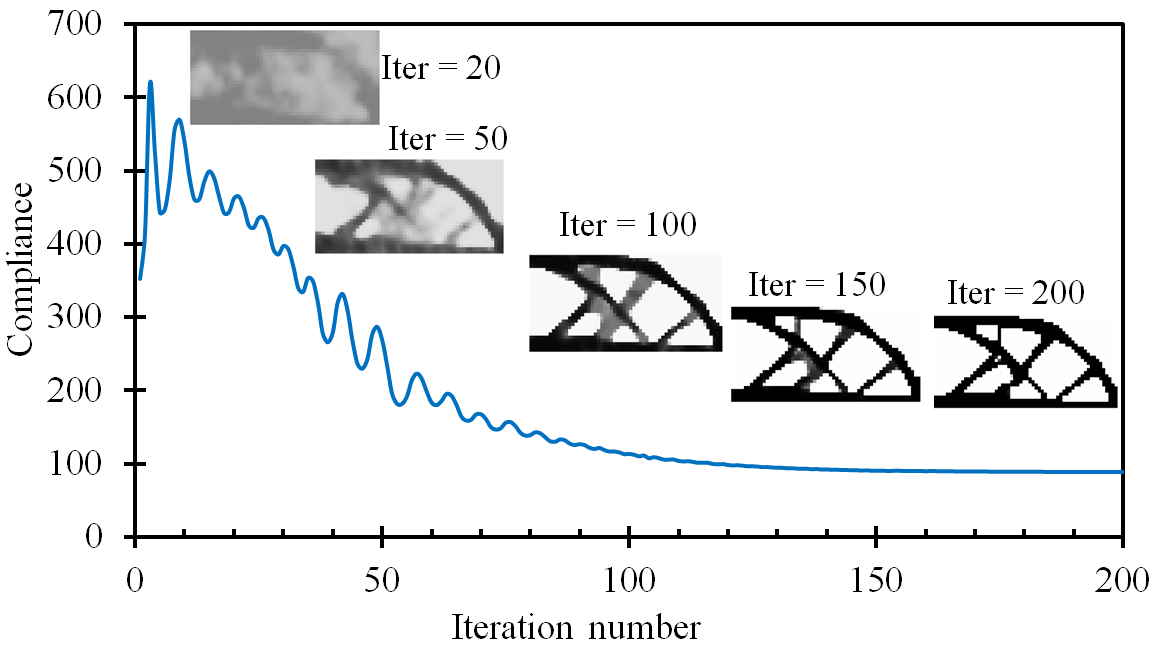}
        \caption*{\small Classical latent vector, Compliance = 88.98}
        \label{fig:loss-quantum}
    \end{subfigure}
    \caption{\small Training loss over iterations using (a) a quantum latent vector (5 qubits, 5 repeating layers), and (b) a classical latent vector sampled from a Gaussian distribution. In both cases, the latent vector is projected to 64 dimensions before decoding.}
    \label{fig:compliance_tip}
\end{figure}

We consider a standard benchmark problem involving a two-dimensional cantilever beam subjected to a downward point load at the free end, as illustrated in Figure~\ref{fig:topology_comparison}a. The cantilever has a length-to-height ratio of \( L/H = 2 \). The left edge is fully fixed, while a unit point load \( F = 1 \) is applied vertically at the bottom corner of the free end. 


\begin{figure}[!tb]
    \centering
    \begin{subfigure}[t]{0.49\textwidth}
        \centering
        \includegraphics[height=4.5cm]{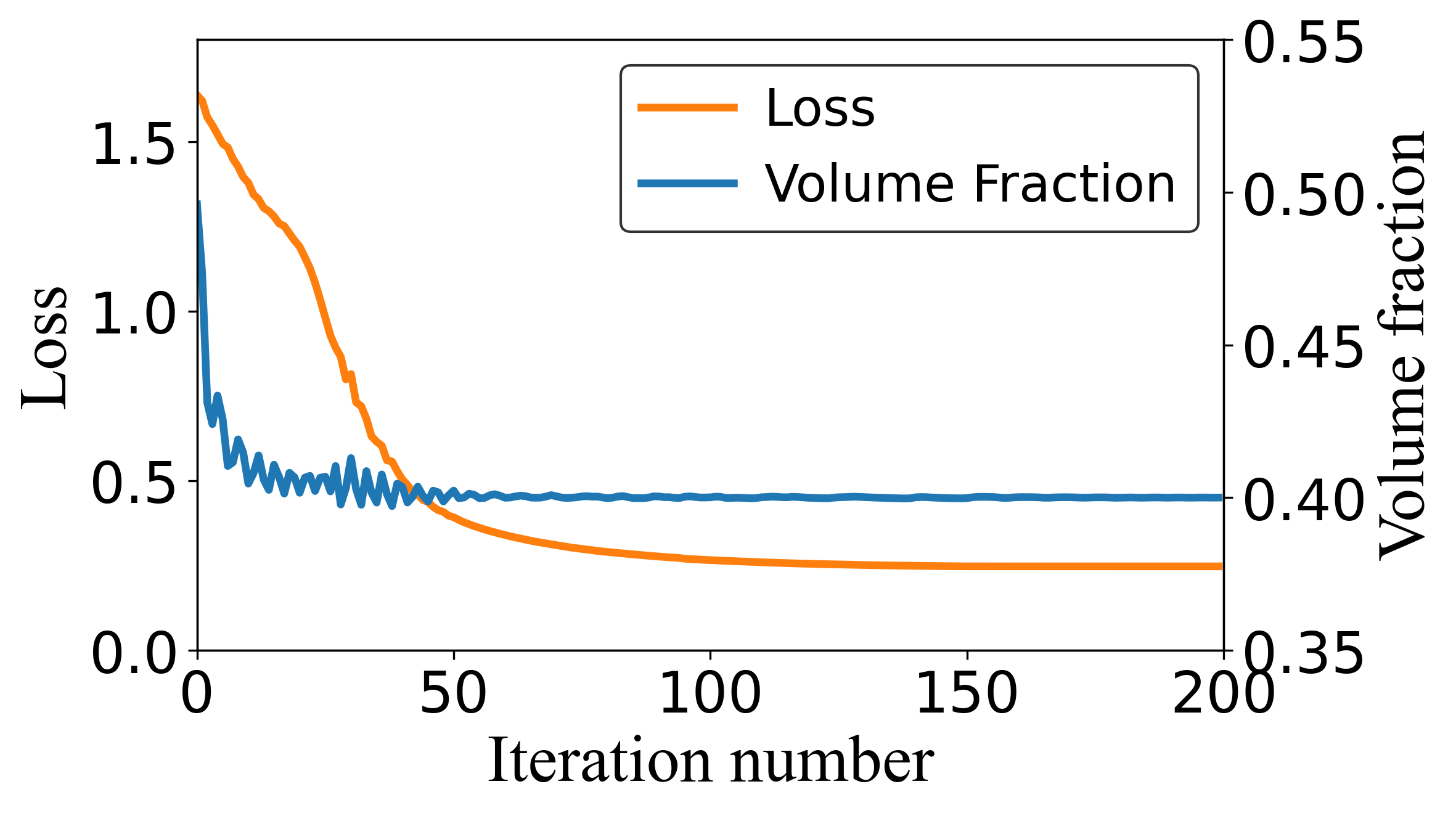}
        \caption*{(a)}
        \label{fig:loss-quantum}
    \end{subfigure}%
    \begin{subfigure}[t]{0.49\textwidth}
        \centering
        \includegraphics[height=4.5cm]{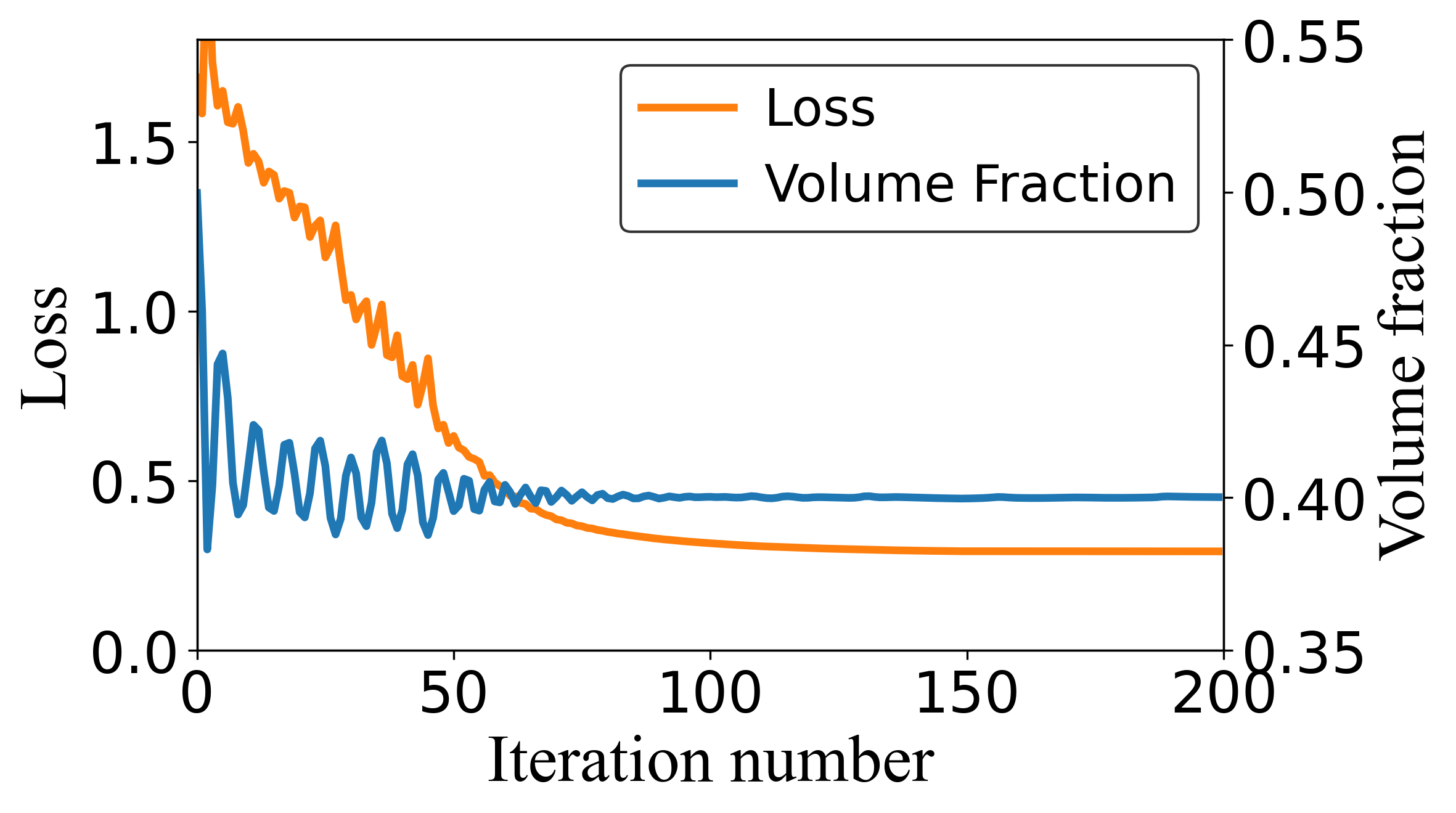}
        \caption*{(b)}
        \label{fig:loss-classical}
    \end{subfigure}
    \caption{Training loss over iterations using (a) a quantum latent vector (5 qubits, 5 layers), and (b) a classical latent vector sampled from a Gaussian distribution. }
    \label{fig:loss-comparison_tip}
\end{figure}


\begin{figure}[!tb]
    \centering
    \begin{subfigure}[t]{0.49\textwidth}
        \centering
        \includegraphics[height=4.5cm, trim=0 0 0 0, clip]{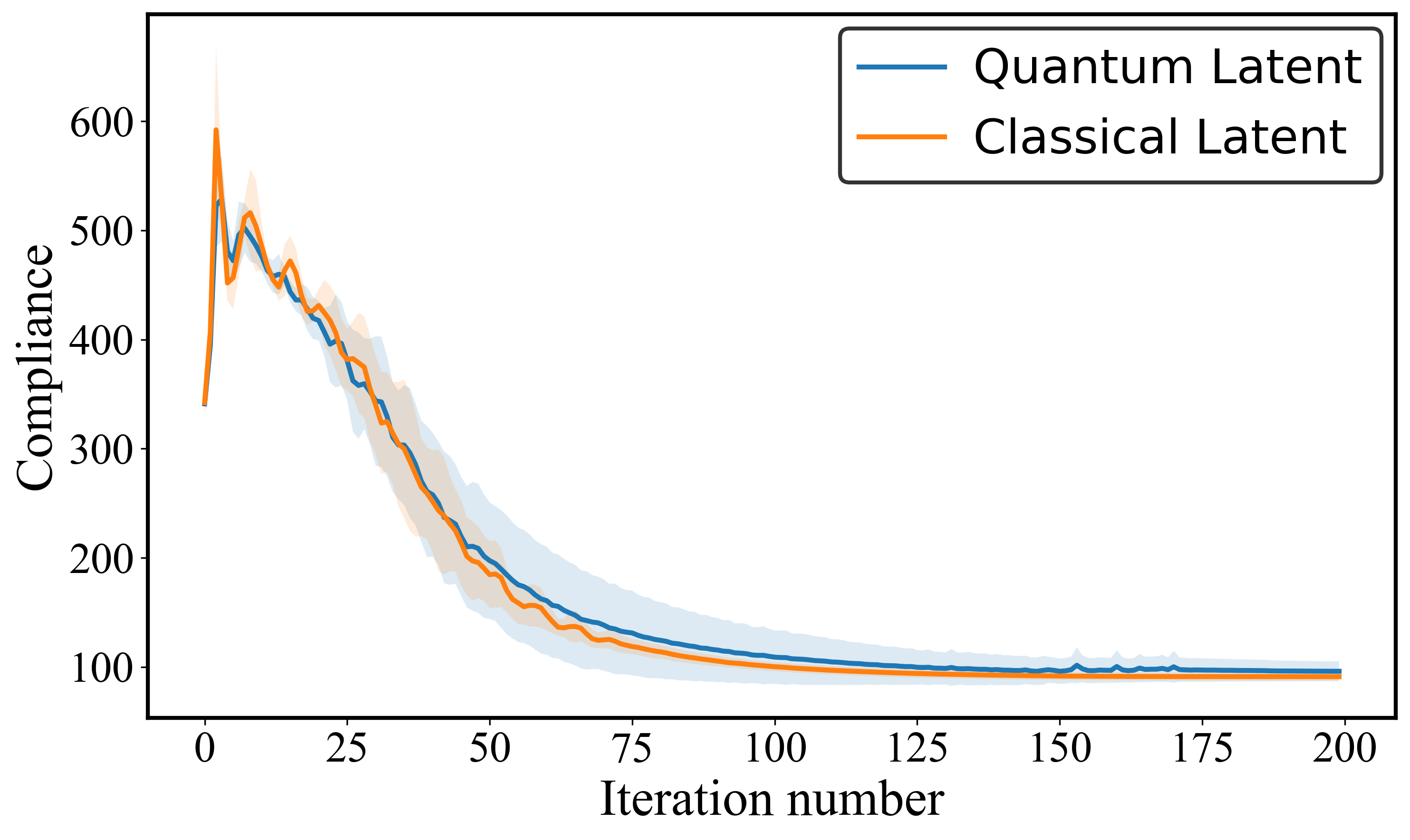}
        \caption*{ (a)}
        \label{fig:loss-quantum}
    \end{subfigure}%
    \begin{subfigure}[t]{0.49\textwidth}
        \centering
        \includegraphics[height=4.5cm]{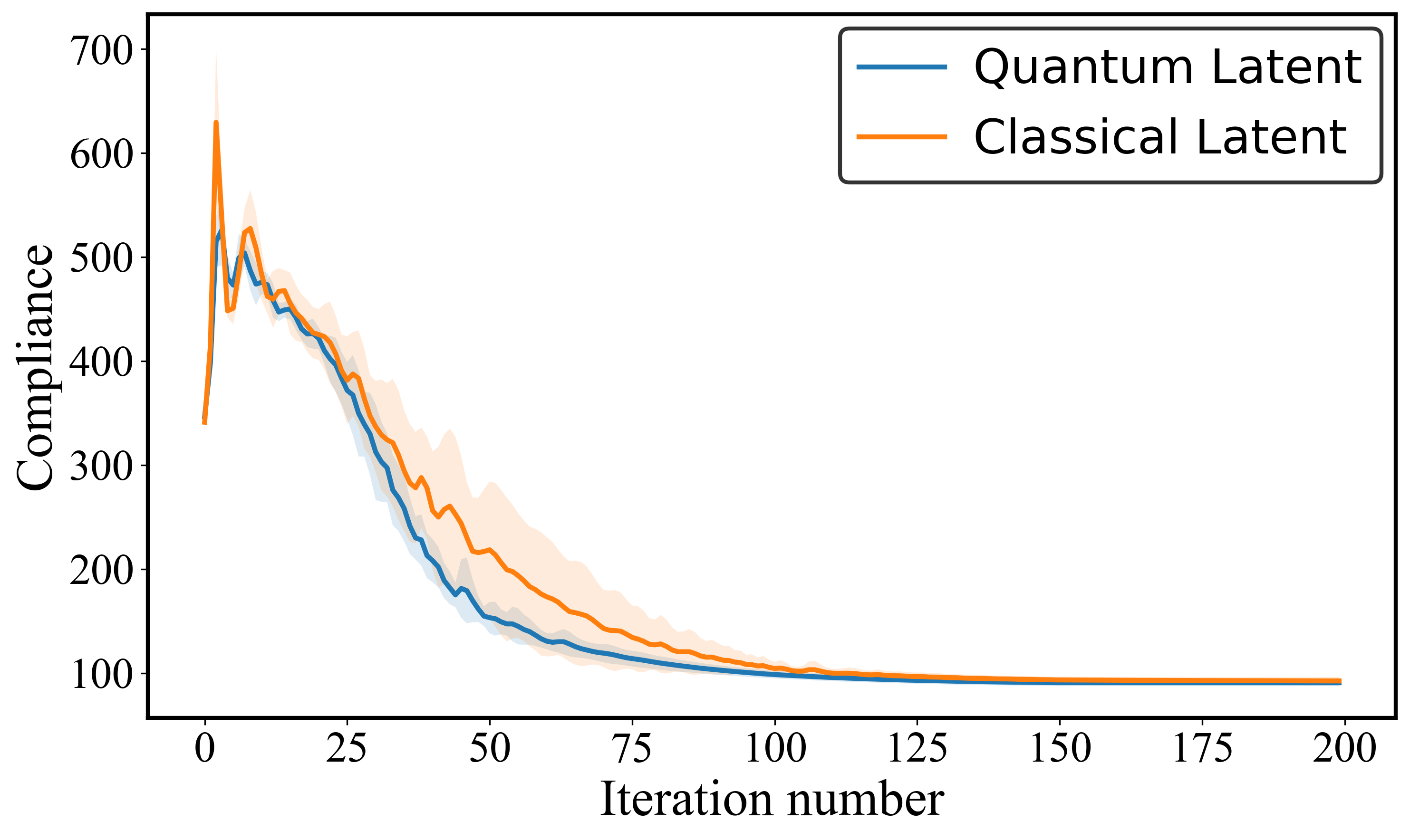}
        \caption*{ (b)}
        \label{fig:loss-classical}
    \end{subfigure}
    
    \vspace{0.5cm}
    
        \begin{subfigure}[t]{0.49\textwidth}
        \centering
        \includegraphics[height=4.5cm]{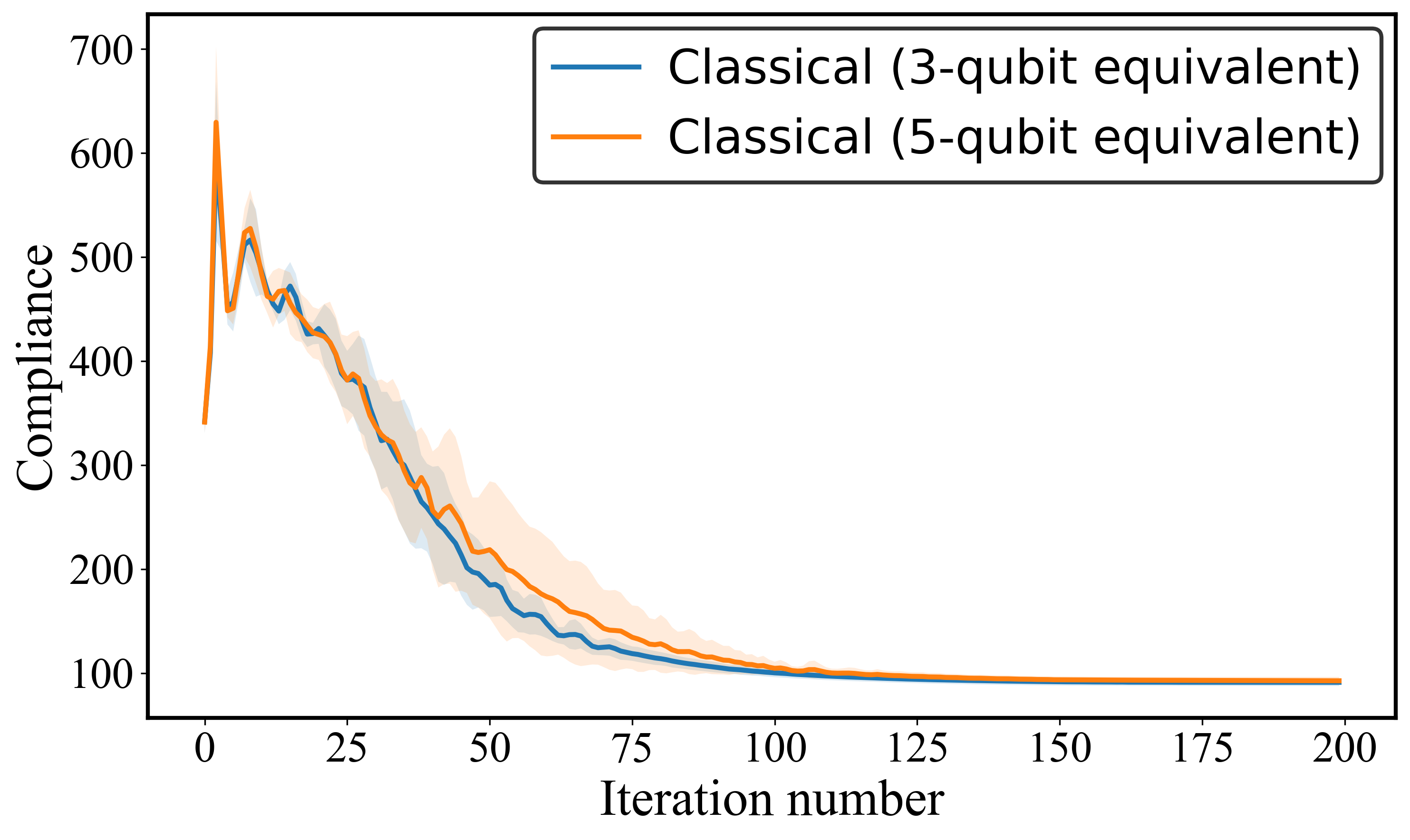}
        \caption*{(c)}
        \label{fig:loss-quantum}
    \end{subfigure}%
    \begin{subfigure}[t]{0.49\textwidth}
        \centering
        \includegraphics[height=4.5cm]{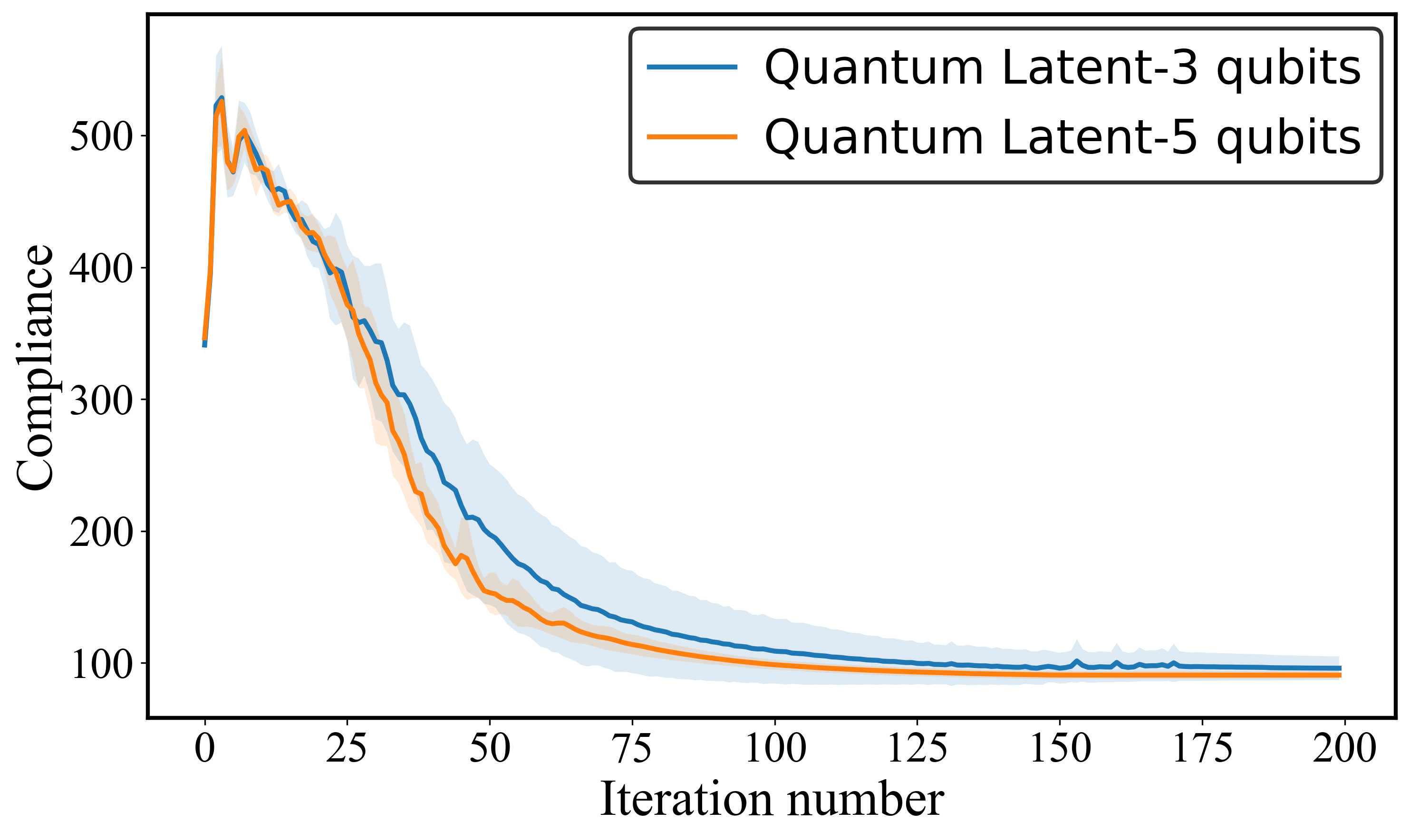}
        \caption*{(d) }
        \label{fig:loss-classical}
    \end{subfigure}
    \caption{Training loss over iterations using (a) a quantum latent vector (5 qubits, 5 layers), and (b) a classical latent vector sampled from a Gaussian distribution. In both cases, the latent vector is projected to 64 dimensions before decoding.}
    \label{fig:loss-comparison_tip1}
\end{figure}



Figure~\ref{fig:topology_comparison} presents three optimized topologies generated using classical and quantum latent vectors generated using latent vectors obtained from a quantum circuit with 5 qubits and 5 entangling layers. Each design corresponds to a different latent vector sampled from the respective encoding scheme, with the associated compliance values reported alongside. These examples demonstrate the variability in structural layouts that can emerge under identical boundary and loading conditions, driven solely by differences in the latent representation. Notably, both classical and quantum encodings yield mechanically sound and interpretable topologies that effectively respond to the prescribed constraints.

Figures~\ref{fig:compliance_tip}a and \ref{fig:compliance_tip}b show the compliance history from representative optimization runs using classical and quantum latent vectors, respectively. Both plots highlight the progressive reduction in compliance over training iterations, reflecting convergence toward an optimal structural layout. To illustrate the evolution of the design during optimization, selected intermediate topologies are overlaid along the timeline. These snapshots demonstrate how the structure becomes increasingly refined, with load-bearing features gradually emerging as the decoder adapts to the underlying latent representation.

Figure~\ref{fig:loss-comparison_tip} compares the evolution of the training loss and the volume fraction for both latent vectors. In each case, the loss function steadily decreases over iterations, while the volume fraction converges toward the target value of 0.4, confirming consistent and stable training behavior for both classical and quantum representations.

To further investigate the impact of quantum circuit size on optimization behavior, we evaluate the performance of 3-qubit quantum encodings using a 5-layer entangling circuit. Figure~\ref{fig:loss-comparison_tip} presents the average compliance trajectories over 10 independent runs for each encoding type—quantum and classical—across both circuit sizes. At each iteration, compliance values are averaged across runs, and the shaded regions denote one standard deviation from the mean, highlighting variability in convergence. The corresponding compliance and design diversity values at iteration 200 are reported in Table~\ref{tab:compliance_diversity_comparison_tip}, providing a quantitative summary of performance across different latent representations.


In Figure~\ref{fig:loss-comparison_tip}a, the 3-qubit quantum encoding is compared with its classical counterpart. Both approaches exhibit a steady reduction in compliance, demonstrating consistent convergence behavior. The convergence trends are similar overall, with the quantum method yielding a slightly higher final compliance. This result suggests that even a compact quantum circuit with only 3 qubits can generate latent representations capable of guiding the decoder toward structurally sound and efficient topologies. 
Figure~\ref{fig:loss-comparison_tip}b shows the comparison to the 5-qubit case. Here, the quantum latent vector achieves faster compliance reduction during the early and intermediate phases of training, outperforming the classical baseline  up to approximately iteration 150. Beyond this point, both methods exhibit stable convergence and reach nearly identical final compliance values.

Subfigures (c) and (d) isolate the effect of latent dimensionality by comparing 3-qubit and 5-qubit encodings within the quantum and classical pipelines, respectively. In the classical case (Figure~\ref{fig:loss-comparison_tip1}(c)), increasing the size of the initial latent vector from 9 to 15 does not noticeably impact the convergence behavior or the final compliance value, indicating limited benefit from additional unstructured latent parameters. In contrast, the quantum results shown in Figure~\ref{fig:loss-comparison_tip1}(d) demonstrate that increasing the circuit size from 3 to 5 qubits—thereby expanding the quantum latent space—leads to faster convergence and consistently lower compliance. This suggests that the additional expressive power provided by a larger quantum circuit enhances the quality of the latent representation and improves overall optimization performance.


\begin{table}[tbp]
\centering
\caption{Mean compliance (± standard deviation) and design diversity of tip loaded cantilever beam at iteration 200 for quantum and classical encodings using 3 and 5 qubits. }
\label{tab:compliance_diversity_comparison_tip}
\begin{tabular}{|c|c|c|c|}
\hline
\textbf{Qubits} & \textbf{Encoding Type} & \textbf{Compliance (mean ± std)} & \textbf{Design Diversity ($L^2$)} \\
\hline
\multirow{2}{*}{3} & Classical & \(91.12 \pm 3.53\) & 125.50 \\
                   & Quantum   & \(94.05 \pm 4.08\) & 136.64 \\
\hline
\multirow{2}{*}{5} & Classical & \(92.60 \pm 3.70\) & 133.44 \\
                   & Quantum   & \(90.83 \pm 3.11\) & 134.33 \\
\hline
\end{tabular}
\end{table}

Table~\ref{tab:compliance_diversity_comparison_tip} also reports the design diversity calculated using \eref{eq:diversity} for the cantilever beam with tip loading. The results indicate that quantum latent vectors consistently yield more diverse topologies than their classical counterparts. In the 3-qubit configuration, the design diversity of the quantum model (136.64) is notably higher than that of the classical encoding (125.50), suggesting a broader exploration of the design space. When the number of qubits is increased to 5, the diversity of the classical model improves (133.44), narrowing the gap with the quantum counterpart (134.33). Nevertheless, quantum encodings maintain a slight advantage across both qubit settings.



\subsection{Simply supported beam with distributed loading on the bottom edge}

\begin{figure}[tb]
    \centering

    \begin{subfigure}[b]{0.45\textwidth}
        \centering
        \includegraphics[width=\textwidth]{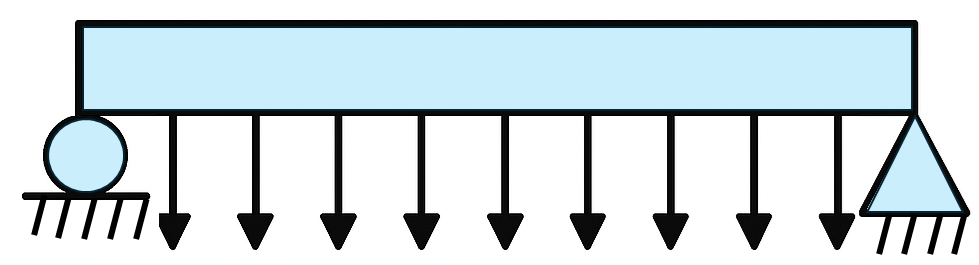}
        \caption{Simply supported beam with distributed loading.}
    \end{subfigure}
    \hfill

    \vspace{0.5cm}

    \begin{subfigure}[b]{0.3\textwidth}
        \centering
        \includegraphics[width=\textwidth]{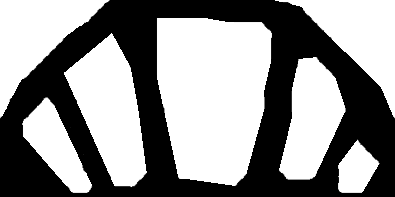}
        \caption{Compliance = 28.72e3}
    \end{subfigure}
    \hfill
    \begin{subfigure}[b]{0.3\textwidth}
        \centering
        \includegraphics[width=\textwidth]{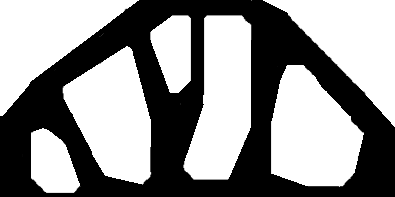}
        \caption{Compliance = 29.91e3}
    \end{subfigure}
    \hfill
    \begin{subfigure}[b]{0.3\textwidth}
        \centering
        \includegraphics[width=\textwidth]{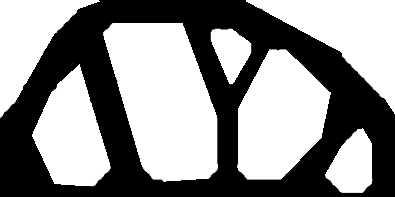}
        \caption{Compliance = 28.22e3}
    \end{subfigure}

    \vspace{0.5cm}

    \begin{subfigure}[b]{0.3\textwidth}
        \centering
        \includegraphics[width=\textwidth]{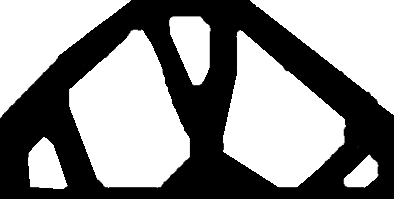}
        \caption{Compliance = 29.16e3}
    \end{subfigure}
    \hfill
    \begin{subfigure}[b]{0.3\textwidth}
        \centering
        \includegraphics[width=\textwidth]{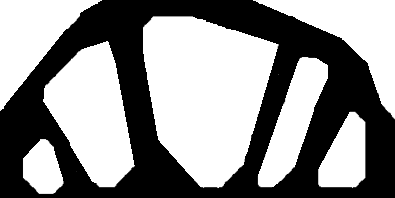}
        \caption{Compliance = 29.36e3}
    \end{subfigure}
    \hfill
    \begin{subfigure}[b]{0.3\textwidth}
        \centering
        \includegraphics[width=\textwidth]{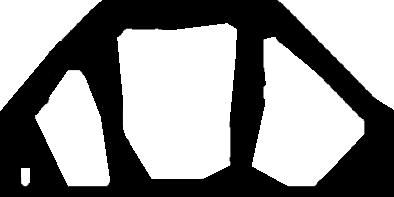}
        \caption{ Compliance = 28.67e3}
    \end{subfigure}

    \caption{Comparison of topologies generated for a simply supported beam using classical (C1–C3) and quantum (Q1–Q3) latent encodings.  Compliance values are listed below each topology.}
    \label{fig:topology_comparison_simpSuppBot1}
\end{figure}

The next benchmark problem, illustrated in Figure~\ref{fig:topology_comparison_simpSuppBot1}a, considers a simply supported beam subjected to a uniformly distributed load of unit magnitude along its bottom edge. The boundary conditions are imposed by restraining the left-bottom corner node in the vertical direction (roller support) and fully constraining the right-bottom corner node in both horizontal and vertical directions (pinned support).
As part of this study, we investigated the effect of increasing the latent space capacity by comparing 3-qubit and 5-qubit quantum circuits. However, only marginal performance gains were observed with 5 qubits, accompanied by increased variance. Given the marginal gains from using additional qubits, and considering the added complexity and potential variability, we focus mainly on 3-qubit encodings for this benchmark.

This experiment focuses on comparing two distinct regularization strategies and their influence on optimization outcomes. The first strategy employs a density projection approach of \eref{proj1}, coupled with loss-based regularization terms—including binary penalization, total variation (TV), and 
$H^1$
  smoothness—to promote physically plausible and well-structured topologies. The second strategy applies spatial density filtering \cite{bendsoe2003topology} followed by a Heaviside projection \cite{guest2004achieving}, enforcing smoothness and discreteness directly at the field level. To isolate the impact of each method, the loss-based regularization terms are disabled when filtering is applied. This controlled comparison enables a detailed evaluation of how each regularization mechanism influences convergence behavior, design quality, and structural interpretability, under otherwise identical modeling conditions and latent encoding schemes.

\begin{figure}[tb]
    \centering

    \begin{subfigure}[b]{0.3\textwidth}
        \centering
        \includegraphics[width=\textwidth]{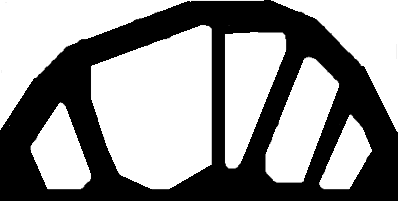}
        \caption{Compliance = 30.00e3}
    \end{subfigure}
    \hfill
    \begin{subfigure}[b]{0.3\textwidth}
        \centering
        \includegraphics[width=\textwidth]{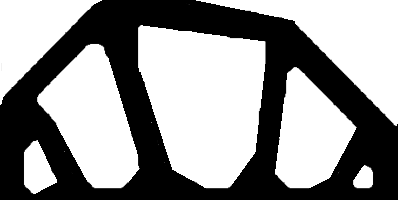}
        \caption{Compliance = 29.65e3}
    \end{subfigure}
    \hfill
    \begin{subfigure}[b]{0.3\textwidth}
        \centering
        \includegraphics[width=\textwidth]{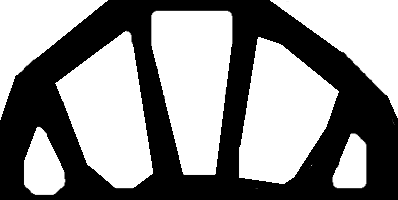}
        \caption{Compliance = 30.02e3}
    \end{subfigure}

    \vspace{0.5cm}

    \begin{subfigure}[b]{0.3\textwidth}
        \centering
        \includegraphics[width=\textwidth]{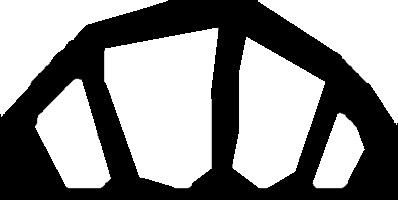}
        \caption{Compliance = 29.32e3}
    \end{subfigure}
    \hfill
    \begin{subfigure}[b]{0.3\textwidth}
        \centering
        \includegraphics[width=\textwidth]{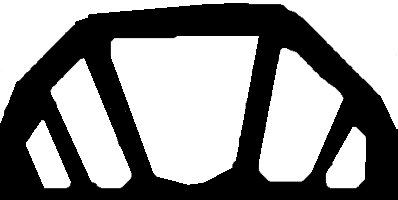}
        \caption{ Compliance = 29.64e3}
    \end{subfigure}
    \hfill
    \begin{subfigure}[b]{0.3\textwidth}
        \centering
        \includegraphics[width=\textwidth]{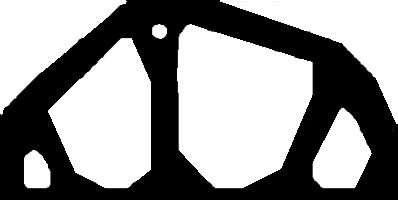}
        \caption{Compliance = 28.88e3}
    \end{subfigure}

    \caption{Comparison of topologies generated for a simply supported beam using filtering and Heaviside projection. }
    \label{fig:topology_comparison_simpSuppBot_filter}
\end{figure}

Selected results from classical and quantum encoding schemes, obtained using both loss-based regularization and traditional density filtering techniques, are shown in Figures~\ref{fig:topology_comparison_simpSuppBot1} and \ref{fig:topology_comparison_simpSuppBot_filter}, respectively. Each topology corresponds to a distinct latent vector sampled from either a 3-qubit quantum circuit or its classical counterpart. Across both regularization strategies, the resulting designs are structurally sound and interpretable, exhibiting comparable compliance values. 
However, notable differences emerge in terms of computational efficiency and convergence behavior. When density filtering and Heaviside projection are employed, the per-iteration computational cost increases substantially compared to loss-based regularization. Moreover, convergence tends to be slower. While loss-regularized training typically reaches high-quality topologies within 200 iterations, filtering-based approaches often require up to 500 iterations to achieve similar results. These observations suggest that, in this setting, loss-based regularization offers a more computationally efficient alternative without sacrificing design quality.

\begin{figure}[tb]
    \centering

    \begin{subfigure}[t]{0.49\textwidth}
        \centering
        \includegraphics[width=\textwidth]{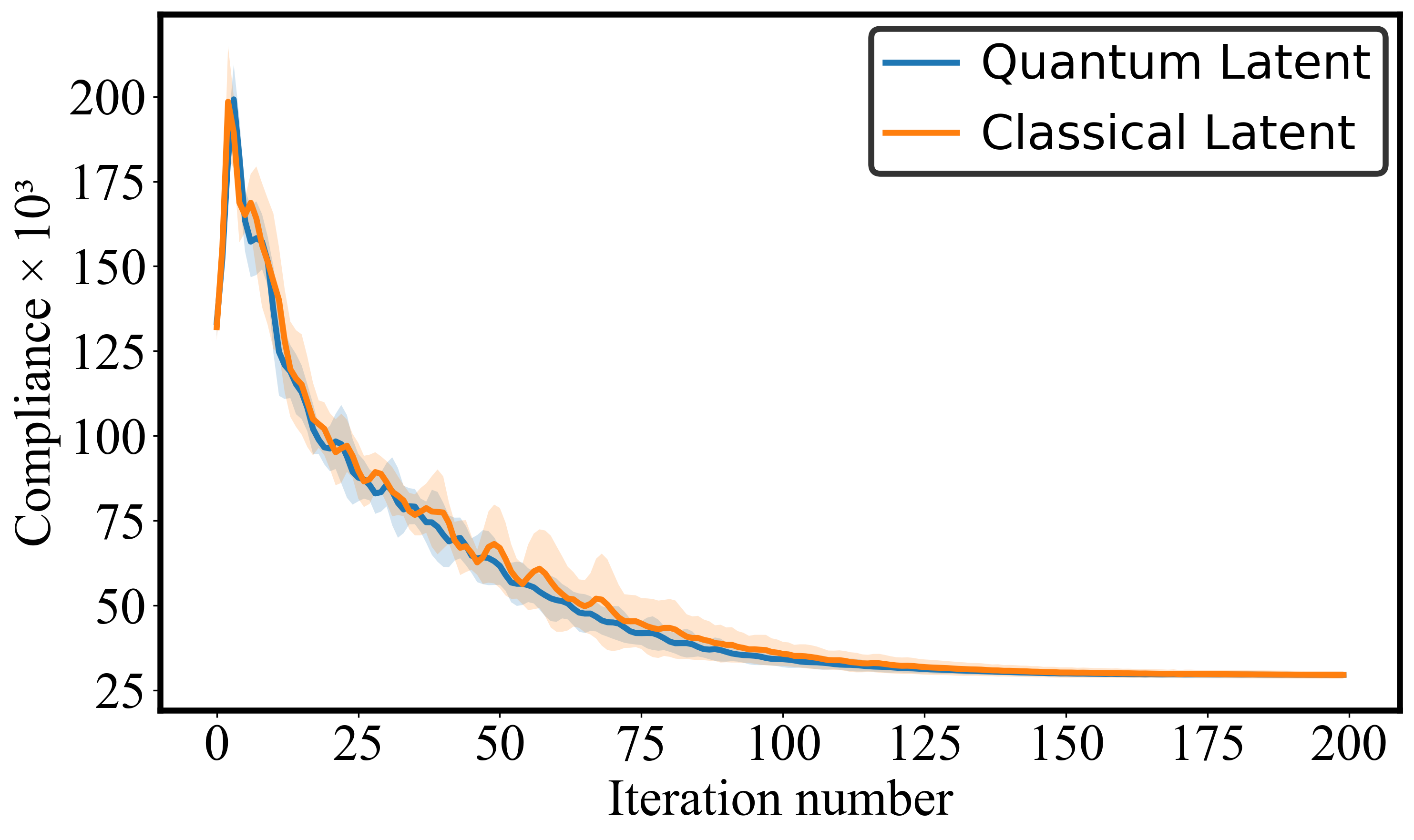}
        \caption{Average compliance over iterations for quantum and classical latent vectors without filtering.}
    \end{subfigure}
    \hfill
    \begin{subfigure}[t]{0.49\textwidth}
        \centering
        \includegraphics[width=\textwidth]{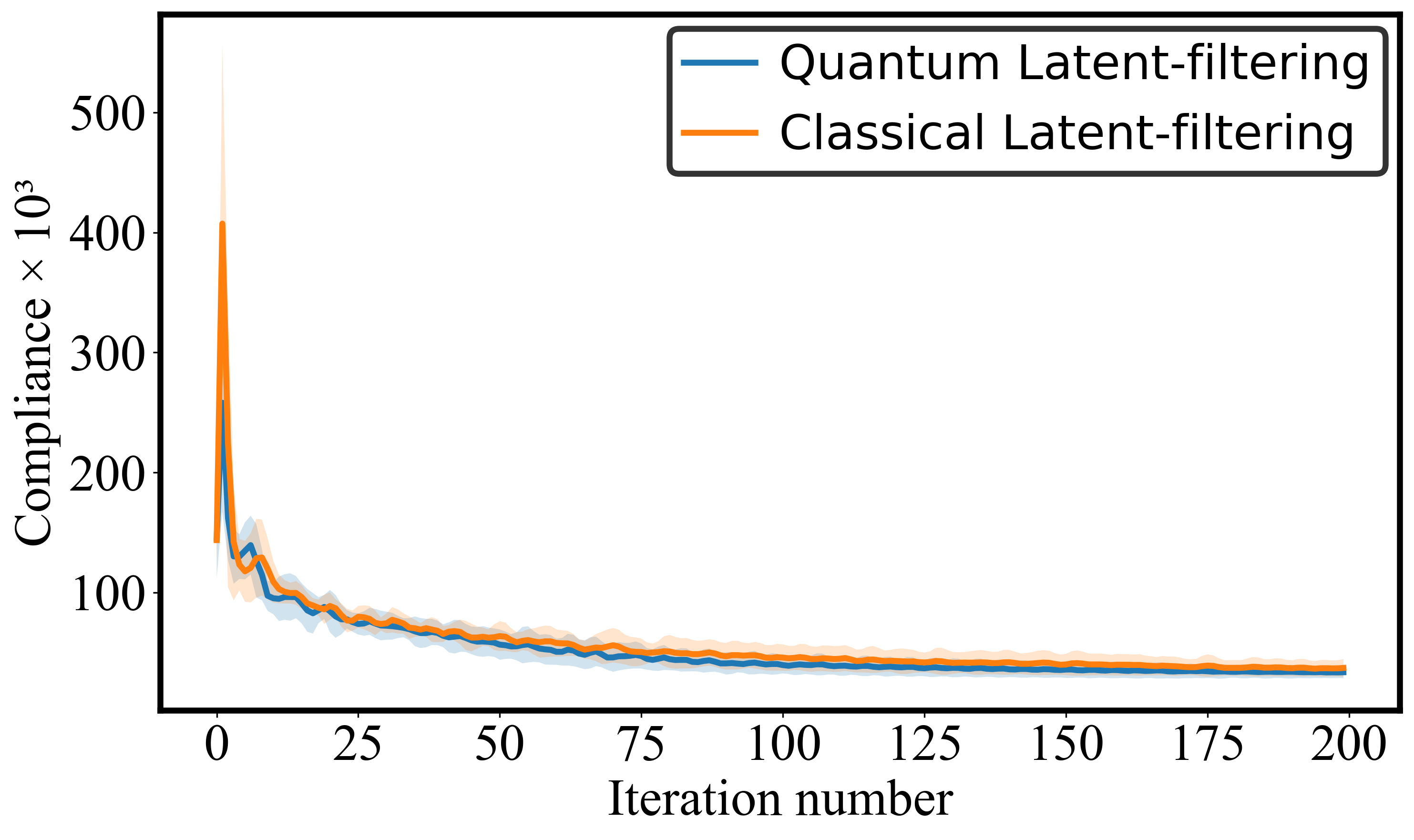}
        \caption{Average compliance over iterations with filtering and Heaviside projection enabled.}
    \end{subfigure}

    \vspace{0.4cm}

    \begin{subfigure}[t]{0.49\textwidth}
        \centering
        \includegraphics[width=\textwidth]{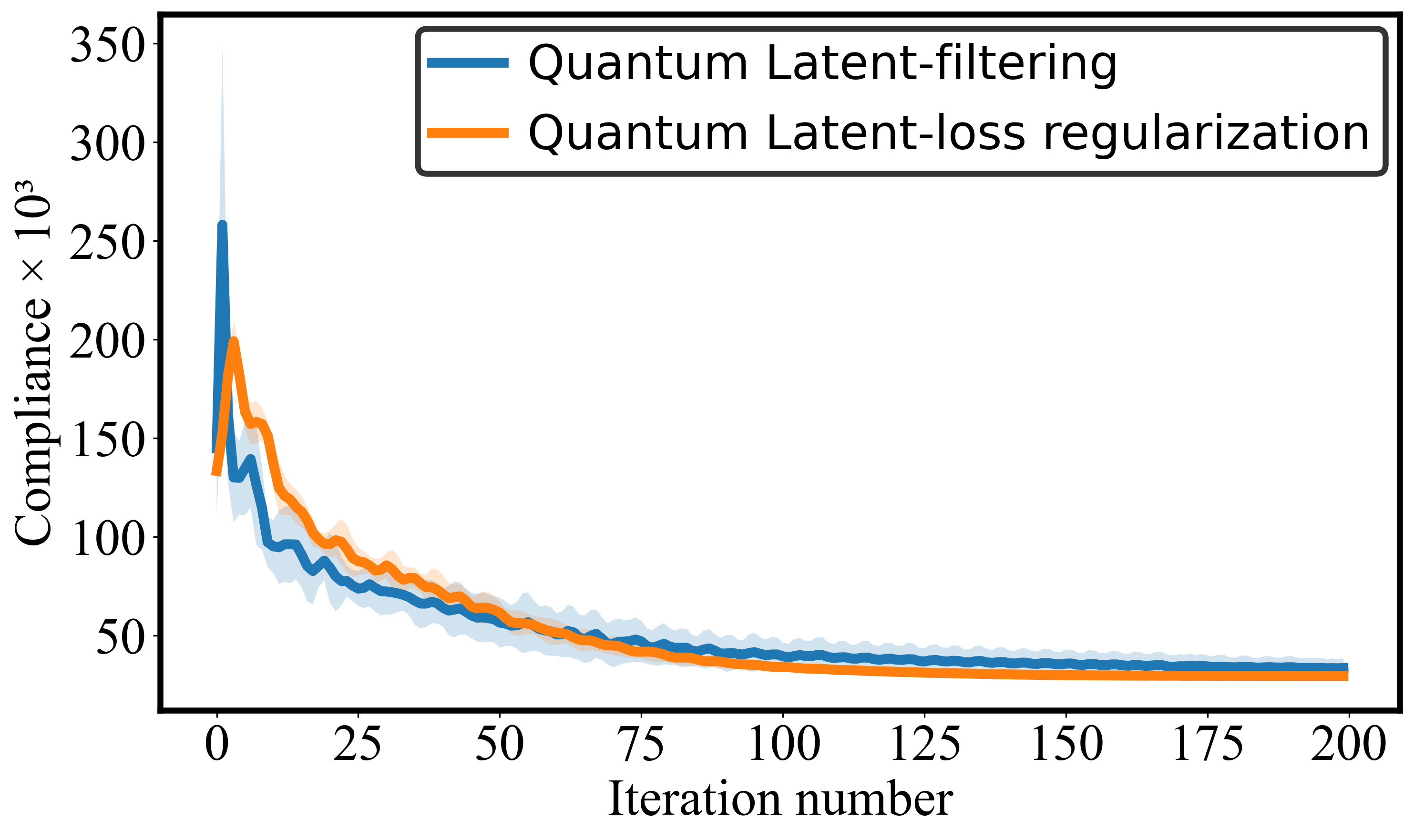}
        \caption{Effect of filtering on compliance using quantum latent vectors.}
    \end{subfigure}
    \hfill
    \begin{subfigure}[t]{0.49\textwidth}
        \centering
        \includegraphics[width=\textwidth]{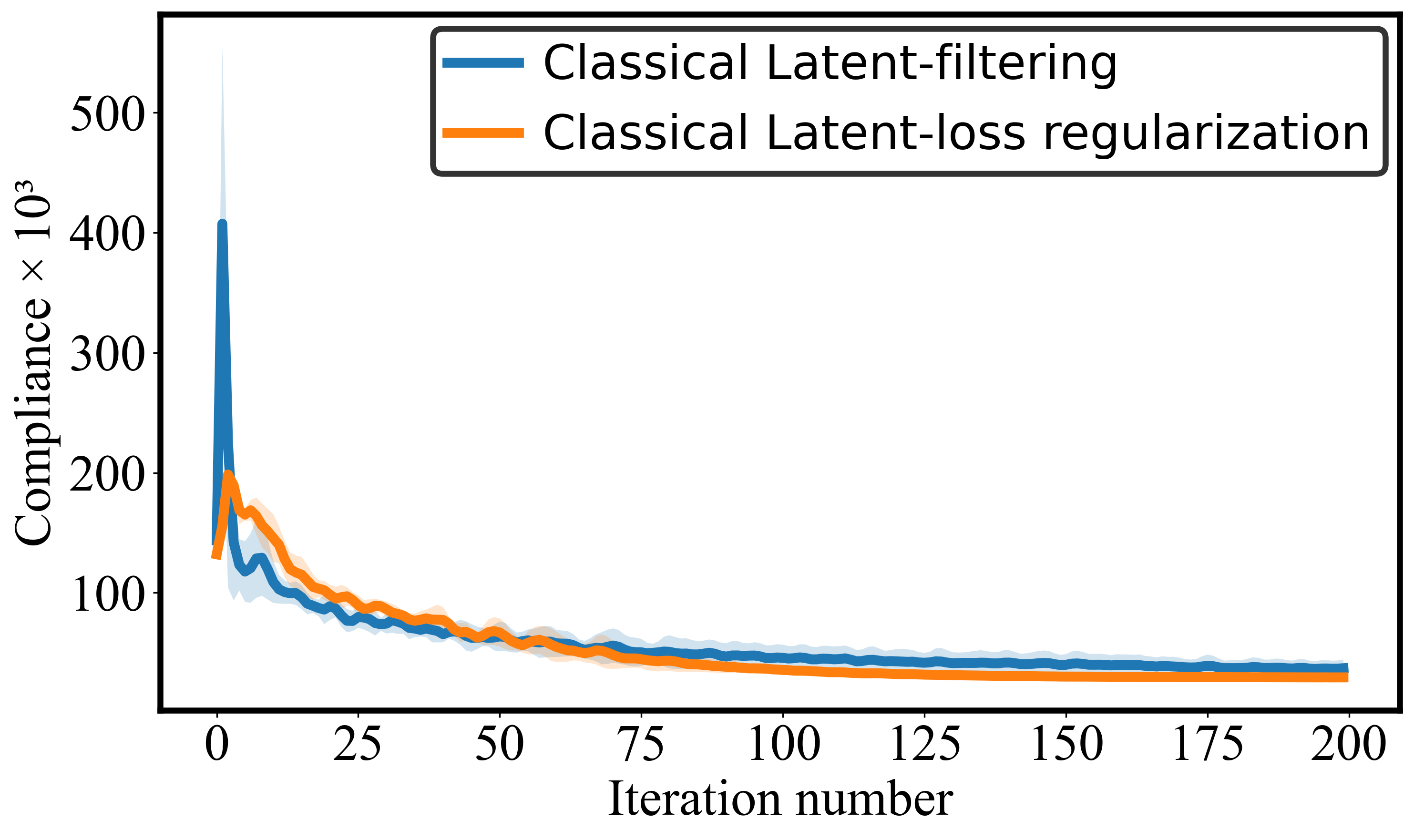}
        \caption{Effect of filtering on compliance using classical latent vectors.}
    \end{subfigure}

    \caption{Comparison of average compliance histories for a simply supported beam under different filtering configurations. Subfigures (a) and (b) show performance with and without filtering, while subfigures (c) and (d) isolate the impact of filtering on quantum and classical latent representations, respectively. Shaded regions represent one standard deviation across ten independent runs.}
    \label{fig:compliance_simpSuppBot}
    
\end{figure}

Figure~\ref{fig:compliance_simpSuppBot} presents the average compliance histories from ten independent optimization runs under various regularization strategies and encoding methods for the simply supported beam benchmark. Each subfigure illustrates the evolution of compliance under a different configuration of filtering and latent encoding, with shaded regions denoting one standard deviation from the mean. Table~\ref{tab:compliance_summary} complements these plots by reporting the mean and standard deviation of compliance at iteration 200, with additional values at iteration 500 for filtered cases—including the results from the 5-qubit configuration.

Across all settings, the variational decoding framework demonstrates stable convergence, with compliance values consistently decreasing over training. When using loss-based regularization, the performance differences between quantum and classical encodings are generally small. For example, in the 3-qubit configuration, the quantum model achieves a slightly lower mean compliance (29.56 vs. 30.07) and marginally reduced variability (0.76 vs. 1.15). Increasing the number of qubits to 5 results in only a modest improvement for the quantum model (29.50 mean compliance), with a slight rise in standard deviation (0.96). In contrast, the classical model performs slightly worse with 5 qubits (30.72 mean, 1.67 std), indicating that unstructured higher-dimensional encodings may introduce variability without delivering consistent gains.


\begin{table}[h!]
\centering
\caption{Compliance and design diversity for the simply supported beam at iteration 200, comparing classical and quantum encodings across different settings. For filtered cases, compliance values at iteration 500 are shown in parentheses. Results are averaged over 10 independent runs.}
\begin{tabular}{|l|l|c|c|c|}
\hline
\textbf{Configuration} & \textbf{Encoding} & \textbf{Compliance} & \textbf{Std. Dev.} & \textbf{Design Diversity (L2)} \\
\hline
\multirow{2}{*}{3 qubits (no filtering)}   
  & Quantum    & 29.56 & 0.76 & 148.46 \\
\cline{2-5}
  & Classical  & 30.07 & 1.15 & 149.23 \\
\hline
\multirow{2}{*}{3 qubits (with filtering)} 
  & Quantum    & 34.75 (30.28) & 5.66 (1.35) & 126.84 \\
\cline{2-5}
  & Classical  & 32.79 (29.98) & 1.96 (0.33) & 125.61 \\
\hline
\multirow{2}{*}{5 qubits (no filtering)}   
  & Quantum    & 29.50 & 0.96 & 149.08 \\
\cline{2-5}
  & Classical  & 30.72 & 1.67 & 151.46 \\
\hline
\end{tabular}
\label{tab:compliance_summary}
\end{table}

\begin{figure}[tb]
    \centering

    \begin{subfigure}[b]{0.45\textwidth}
        \centering
        \includegraphics[width=\textwidth]{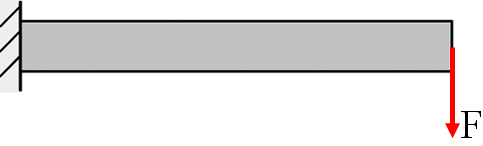}
        \caption{Simply supported beam with distributed loading.}
    \end{subfigure}
    \hfill

\vspace{0.5cm}

    \begin{subfigure}[b]{0.3\textwidth}
        \centering
        \includegraphics[width=\textwidth]{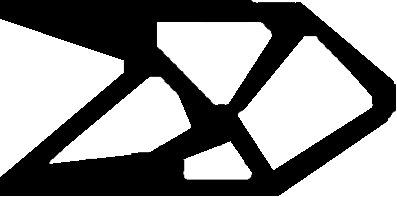}
        \caption{Compliance = 81.96}
    \end{subfigure}
    \hfill
    \begin{subfigure}[b]{0.3\textwidth}
        \centering
        \includegraphics[width=\textwidth]{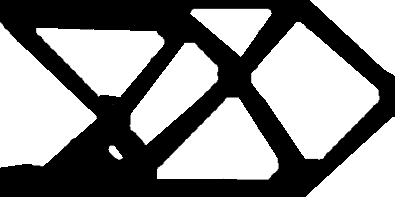}
        \caption{Compliance = 84.41}
    \end{subfigure}
    \hfill
    \begin{subfigure}[b]{0.3\textwidth}
        \centering
        \includegraphics[width=\textwidth]{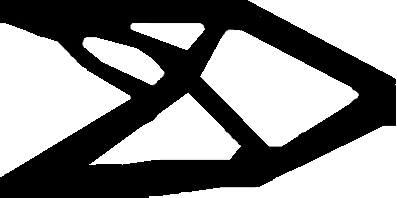}
        \caption{Compliance = 80.80}
    \end{subfigure}

    \vspace{0.5cm}

    \begin{subfigure}[b]{0.3\textwidth}
        \centering
        \includegraphics[width=\textwidth]{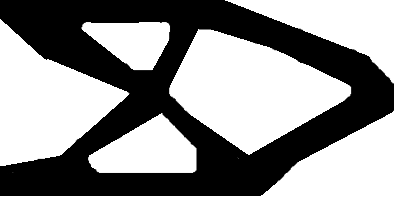}
        \caption{Compliance = 82.10}
    \end{subfigure}
    \hfill
    \begin{subfigure}[b]{0.3\textwidth}
        \centering
        \includegraphics[width=\textwidth]{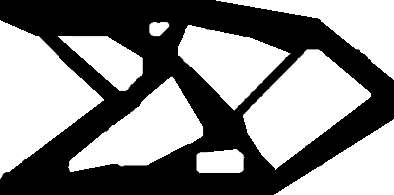}
        \caption{Compliance = 81.73}
    \end{subfigure}
    \hfill
    \begin{subfigure}[b]{0.3\textwidth}
        \centering
        \includegraphics[width=\textwidth]{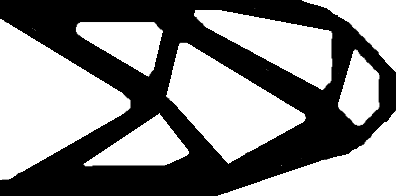}
        \caption{Compliance = 85.55}
    \end{subfigure}

\caption{Comparison of optimized topologies for a mid-point loaded cantilever beam (geometry shown in (a)) using classical (b--d) and quantum (e--g) latent encodings. All results are obtained using loss-based regularization, including volume, smoothness, and discreteness penalties.}
    \label{fig:topology_comparison_cant_tip   }
\end{figure}

The more significant distinction emerges between the two regularization strategies. Traditional density filtering and Heaviside projection, while capable of producing high-quality topologies, require considerably more iterations and computational resources to converge. As shown in subfigures~(b)–(d), filtered models converge more slowly and consistently yield higher compliance values in early and mid-stage iterations. At iteration 200, the quantum filtered model has a compliance of 34.75 with high variability (std = 5.66), only approaching competitive values (30.28) after 500 iterations. The classical filtered model performs more stably (32.79 at iteration 200; 29.98 at 500), but still lags behind the loss-regularized counterparts in efficiency.

\subsection{Cantilever beam subjected to a point load at the mid-height of the free edge}
This benchmark problem considers a cantilever beam subjected to a concentrated vertical load applied at the midpoint of its free (right) edge. Unlike the tip-loaded case, this loading configuration is symmetric about the horizontal axis, resulting in a different stress distribution and encouraging the formation of alternative structural pathways. Owing to the symmetric nature of the applied load and boundary conditions, this example also provides an opportunity to examine whether the decoder, when driven by different latent encodings, can preserve geometric symmetry in the optimized designs.

To investigate this, we first performed topology optimization without applying any explicit symmetry constraints. The results, shown in Figure~\ref{fig:topology_comparison_cant_tip }, reveal that the generated topologies often deviate from symmetry, despite the symmetric problem setup. This suggests that the decoder, when driven purely by latent vectors and guided only by physical loss terms, does not inherently enforce geometric symmetry—especially in the absence of architectural or input-based symmetry bias.

\begin{figure}[tb]
    \centering
    \caption*{Classical latent vectors}
    \begin{subfigure}[b]{0.3\textwidth}
        \centering        \includegraphics[width=\textwidth]{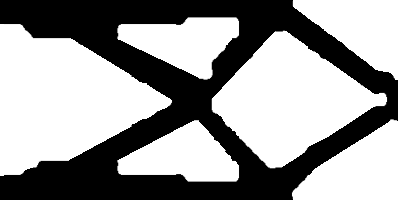}
        \caption{Compliance = 87.10}
    \end{subfigure}
    \hspace{0.03\textwidth}
    \begin{subfigure}[b]{0.3\textwidth}
        \centering        \includegraphics[width=\textwidth]{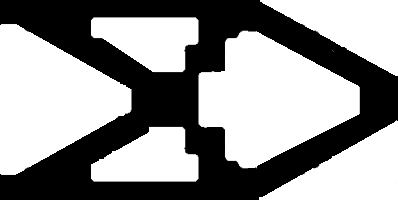}
        \caption{Compliance = 89.95}
    \end{subfigure}
    \hspace{0.03\textwidth}
    \begin{subfigure}[b]{0.3\textwidth}
        \centering        \includegraphics[width=\textwidth]{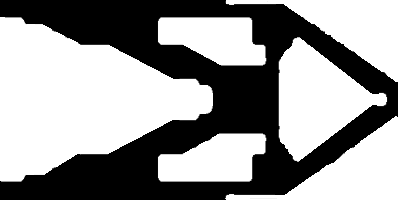}
        \caption{Compliance = 89.88}
    \end{subfigure}
    \\[0.5cm]  
    \caption*{Quantum latent vectors}
    \begin{subfigure}[b]{0.3\textwidth}
        \centering        \includegraphics[width=\textwidth]{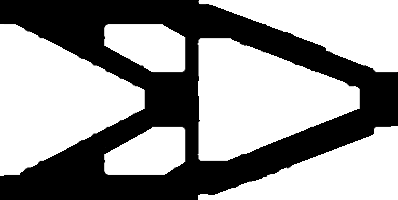}
        \caption{Compliance = 88.28}
    \end{subfigure}
    \hspace{0.03\textwidth}
    \begin{subfigure}[b]{0.3\textwidth}
        \centering        \includegraphics[width=\textwidth]{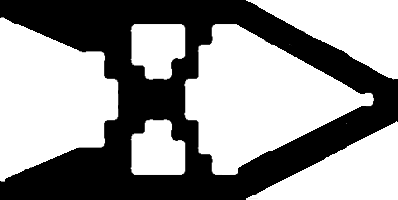}
        \caption{Compliance = 89.13}
    \end{subfigure}
    \hspace{0.03\textwidth}
    \begin{subfigure}[b]{0.3\textwidth}
        \centering        \includegraphics[width=\textwidth]{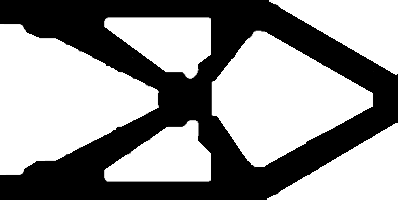}
        \caption{Compliance = 89.13}
    \end{subfigure}
    \caption{Comparison of optimized topologies for the cantilever beam with midpoint loading, generated using classical and quantum latent vectors. The designs are obtained using standard density filtering combined with the Heaviside projection. The top row shows results from classical latent vectors, while the bottom row shows results from quantum latent vectors.}    \label{fig:topology_comparison_cant_tip_sym}
\end{figure}

To address the lack of symmetry in the optimized results, several strategies for promoting symmetry during training can be considered. One approach involves hard enforcement by mirroring the decoded density field about the horizontal axis. While this guarantees symmetry, it restricts the decoder’s ability to explore diverse structural configurations. Alternatively, a soft regularization strategy is adopted by introducing a penalty term that discourages deviations between the density field and its horizontally mirrored counterpart. This symmetry term is defined as
\begin{equation}
\mathcal{L}_{\text{sym}} = \lambda_{\text{sym}} \sum_{x, y} \left( \rho(x, y) - \rho(x, -y) \right)^2,
\end{equation}
where \( \lambda_{\text{sym}} \) controls the strength of the regularization. To avoid overly constraining the optimization in its early stages, a continuation strategy is employed in which \( \lambda_{\text{sym}} \) increases linearly over the course of training. This approach allows broad exploration of the design space initially, while gradually encouraging symmetry as the optimization progresses. The resulting topologies, shown in Figure~\ref{fig:topology_comparison_cant_tip_sym}, exhibit significantly improved symmetry while maintaining competitive compliance values, demonstrating the effectiveness of the soft symmetry constraint.

\begin{figure}[tb]
    \centering

    \begin{subfigure}[t]{0.49\textwidth}
        \centering
        \includegraphics[width=\textwidth]{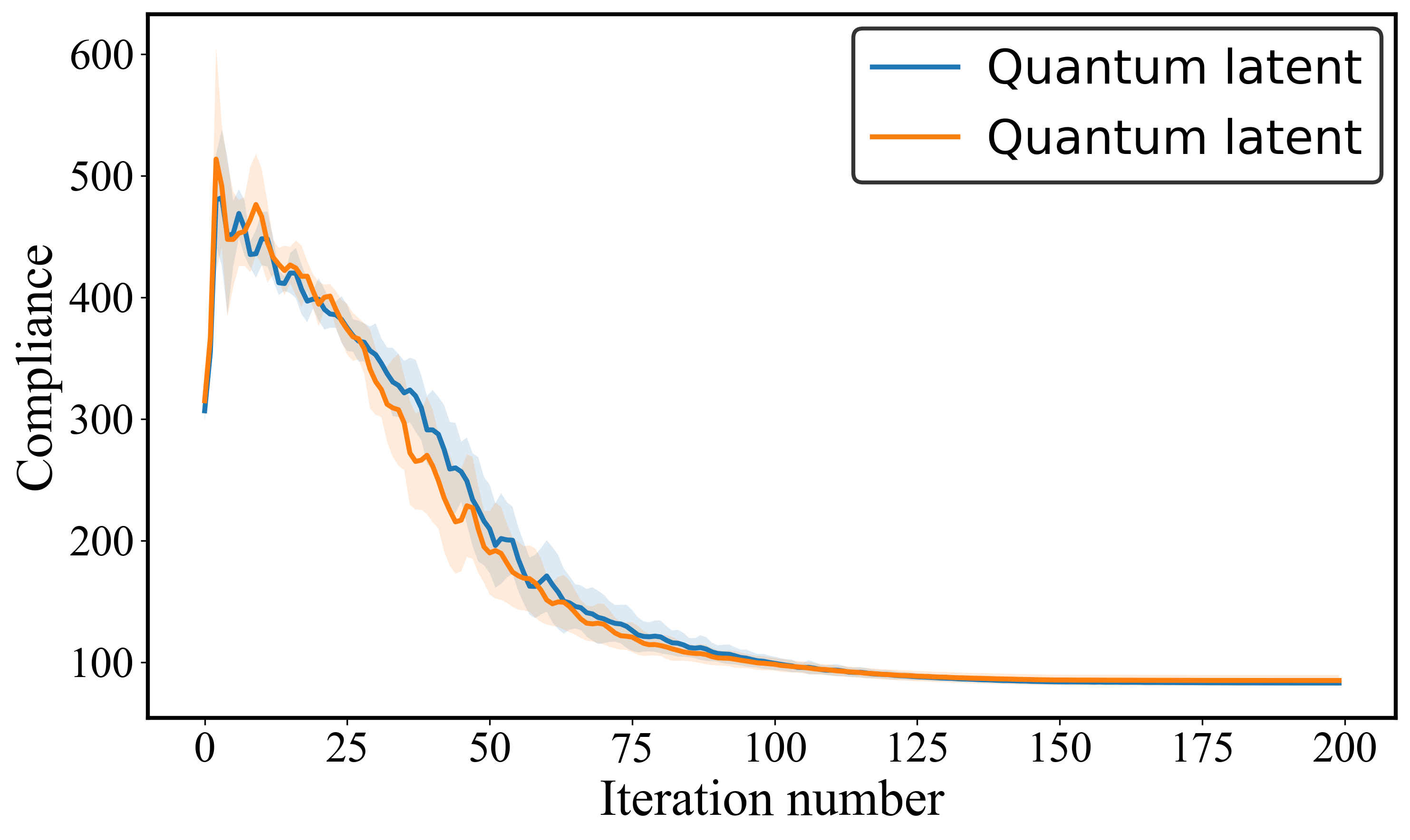}
        \caption{Average compliance over iterations without symmetry enforcement.}
    \end{subfigure}
    \hfill
    \begin{subfigure}[t]{0.49\textwidth}
        \centering
        \includegraphics[width=\textwidth]{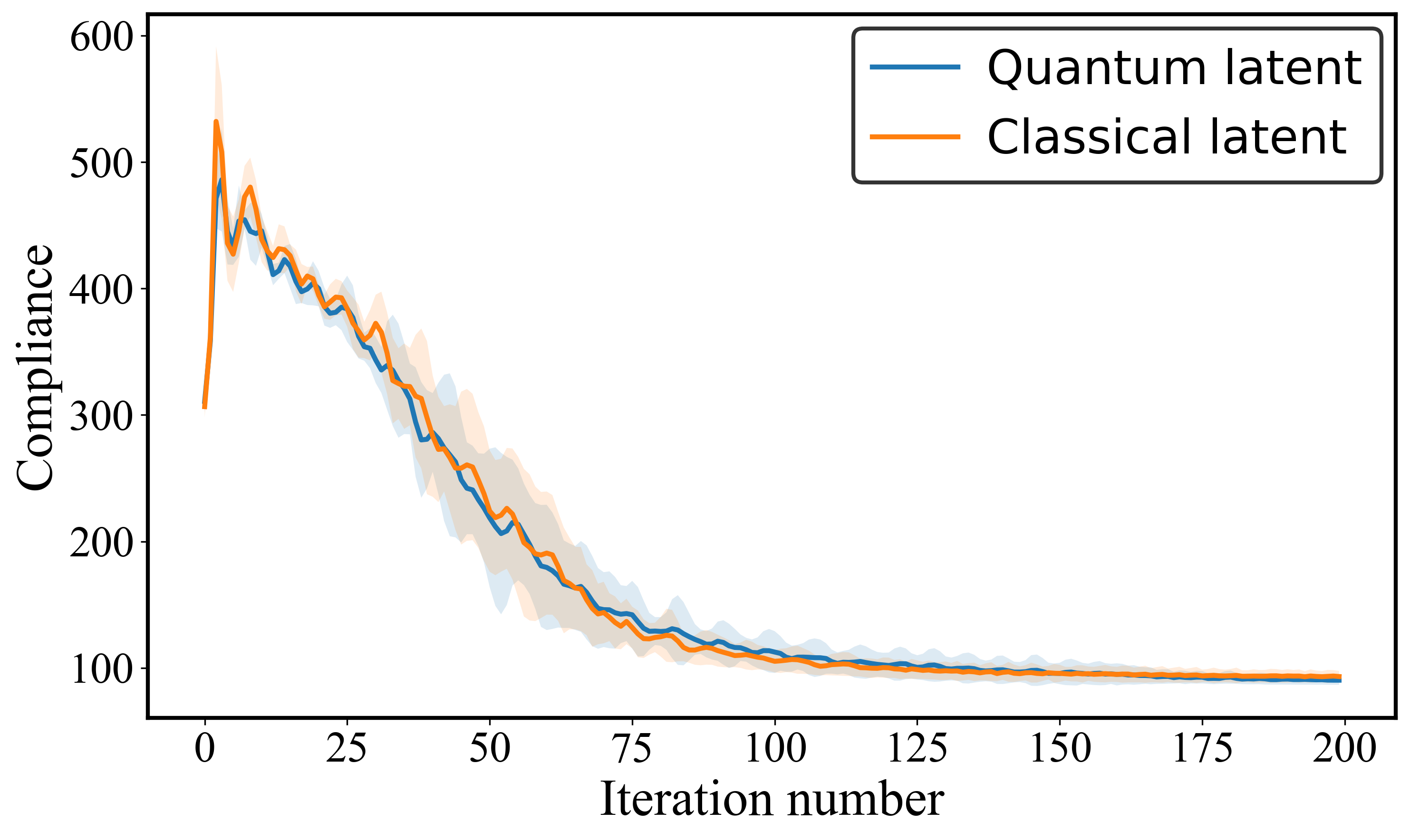}
        \caption{Average compliance over iterations with symmetry regularization.}
    \end{subfigure}

    \vspace{0.4cm}

    \begin{subfigure}[t]{0.49\textwidth}
        \centering
        \includegraphics[width=\textwidth]{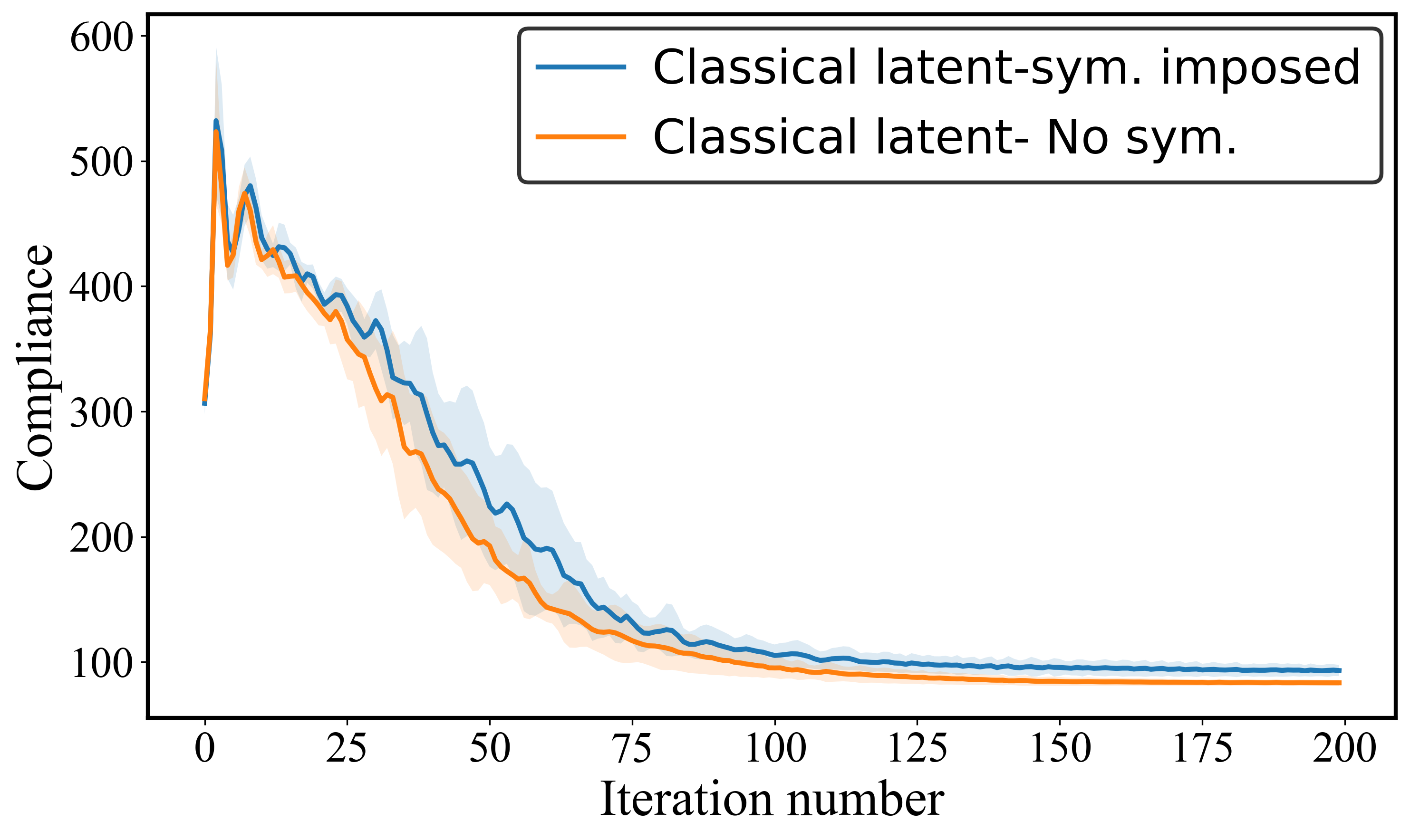}
        \caption{Comparison of average compliance using classical latent vectors, with and without symmetry.}
    \end{subfigure}
    \hfill
    \begin{subfigure}[t]{0.49\textwidth}
        \centering
        \includegraphics[width=\textwidth]{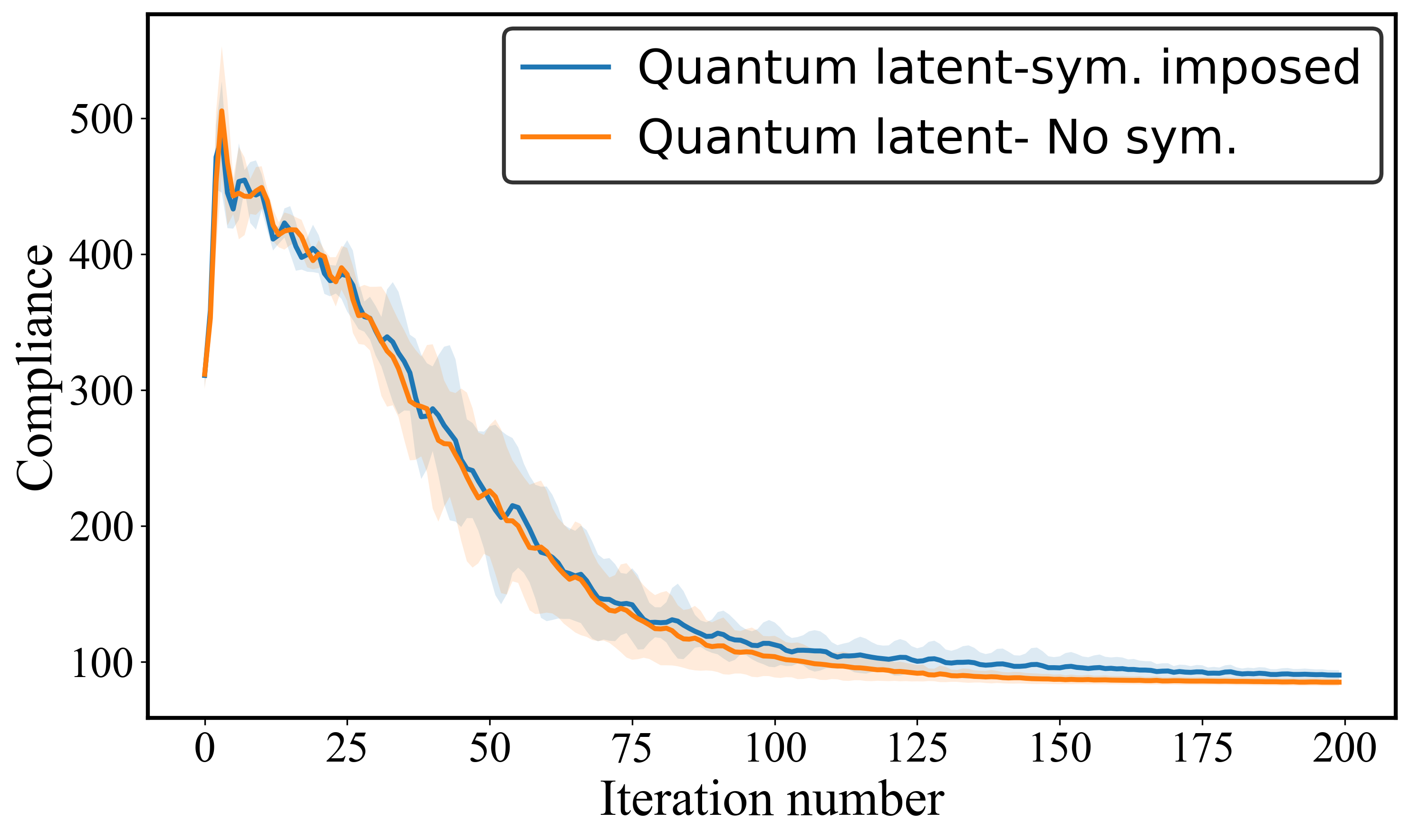}
        \caption{Comparison of average compliance using quantum latent vectors, with and without symmetry.}
    \end{subfigure}

    \caption{Average compliance histories for the cantilever beam with midpoint tip loading. Subfigures (a) and (b) show the effect of not enforcing and enforcing symmetry, respectively. Subfigures (c) and (d) provide direct comparisons of symmetry vs.\ no symmetry for classical and quantum encodings.}
    \label{fig:avg_compliance_symmetry_comparison}
\end{figure}

Figure~\ref{fig:avg_compliance_symmetry_comparison} and Table~\ref{tab:mid_cantilever_compliance_diversity} present a comprehensive comparison of compliance histories and statistical outcomes. Various combinations of latent encoding (quantum vs.\ classical), qubit capacity (3 vs.\ 5), and symmetry regularization (imposed vs.\ not imposed) are analyzed to understand the impact of these factors on structural performance and convergence behavior.

Subfigure~(a) shows the average compliance evolution without symmetry enforcement. Both quantum and classical latent vectors result in consistent convergence, with the quantum encoding yielding slightly lower final compliance and reduced standard deviation, indicating improved consistency across runs. These results serve as a baseline for evaluating the effects of symmetry regularization.

Subfigure~(b) illustrates the compliance trends when symmetry is imposed via soft regularization. Compliance values are generally higher in this case, particularly during early and intermediate stages of training. Increased variance is also observed, which suggests that enforcing symmetry initially restricts the decoder’s flexibility but eventually guides the topology toward symmetric, physically interpretable layouts. The contrast with subfigure~(a) highlights the trade-off between geometric regularity and structural efficiency.

Subfigures~(c) and (d) compare compliance trends for classical and quantum encodings, respectively, with and without symmetry. Subfigure~(c) shows that classical latent vectors experience a noticeable increase in both compliance and variability when symmetry is imposed. This sensitivity reflects the unstructured nature of classical encodings, which are less adaptable under constrained optimization. In contrast, subfigure~(d) demonstrates that quantum encodings maintain more stable convergence even with symmetry regularization. The increase in compliance is modest, and variability remains relatively low, indicating that the quantum latent space more naturally accommodates symmetry constraints.

Table~\ref{tab:mid_cantilever_compliance_diversity} presents the design diversity for the cantilever beam with midpoint loading, measured as the mean pairwise \( L^2 \) distance between optimized topologies across 10 independent runs. In the absence of symmetry enforcement, both quantum and classical encodings exhibit high diversity, with quantum encodings showing slightly higher variability in the 5-qubit case (142.40 vs.\ 130.60), but slightly lower in the 3-qubit case (141.64 vs.\ 144.28). These values suggest that both encoding types are capable of exploring a rich set of structural layouts, with quantum encodings generally maintaining a broader design space as the latent dimension increases.

When symmetry is enforced via regularization, design diversity consistently decreases across all configurations. This reduction is expected, as symmetry constraints limit the number of admissible solutions. The drop is especially noticeable in the 3-qubit case, where diversity falls from 144.28 to 112.16 for classical encoding, and from 141.64 to 111.46 for quantum encoding. Similar trends are observed in the 5-qubit case, with quantum and classical diversity reduced to 118.56 and 117.11, respectively. Notably, even under symmetry constraints, quantum encodings retain slightly higher diversity, suggesting a greater capacity to express a range of symmetric, yet mechanically distinct, topologies.

Overall, these results highlight a clear trade-off between structural diversity and geometric regularity. Symmetry constraints improve interpretability and visual consistency of the designs but reduce variability. Quantum encodings, particularly in the higher qubit setting, appear to offer a favorable balance between diversity and constraint satisfaction.

\section{Conclusion}
This work presented a variational decoding framework for topology optimization that leverages both classical and quantum latent encodings. By projecting low-dimensional latent vectors—sampled from classical distributions or generated via parameterized quantum circuits—into a higher-dimensional design space, we enable resolution-independent structure generation through a coordinate-based neural decoder. The decoder is trained end-to-end using physics-informed loss functions, allowing gradient-based optimization of either classical parameters or quantum circuit weights.

Through systematic comparisons, it was demonstrated that quantum encodings, even with as few as 3–5 qubits, are capable of generating diverse and high-quality topologies. In several cases, quantum latents exhibited faster convergence and lower variance compared to classical baselines. Additionally, we analyzed the effects of circuit depth and latent dimensionality on performance, showing that quantum encodings can benefit from increased qubit count without requiring higher projection dimensionality. 
 Different regularization strategies, including sharp density projections with explicit smoothness penalties and traditional filtering-based methods, were also evaluated. Results suggest that both approaches can yield discrete and interpretable topologies, but they offer distinct trade-offs in convergence speed, gradient flow, and computational cost.

Overall, this work offers a hybrid framework that combines the flexibility of neural decoders with the expressive capacity of quantum circuits. It opens promising avenues for further exploration, including reinforcement learning-based latent optimization, geometry-conditioned quantum encodings, and quantum-classical co-design strategies for structural design tasks.

\section{Acknowledgment}
 This work was supported by the Institute of Digital Engineering (IDE). The author gratefully acknowledges the funding provided by IDE, which made this research possible.

\begin{table}[!tb]
\centering
\caption{Compliance and design diversity for the cantilever beam with midpoint loading at iteration 200, grouped by qubit number and symmetry condition. Values are averaged over 10 independent runs.}
\begin{tabular}{|c|c|c|c|c|}
\hline
\textbf{Qubits} & \textbf{Latent type} & \textbf{Symmetry} & \textbf{Compliance (mean ± std)} & \textbf{Design Diversity ($L^2$)} \\
\hline
\multirow{2}{*}{3} & Quantum   & No  & 82.83 ± 1.81  & 141.64 \\

                   & Classical & No  & 84.87 ± 4.50  & 144.28 \\
\hline
\multirow{2}{*}{5} & Quantum   & No  & 83.07 ± 2.77  & 142.40 \\
                   & Classical & No  & 83.42 ± 2.36  & 130.60 \\
\hline
\multirow{2}{*}{3} & Quantum   & Yes & 95.63 ± 10.64 & 111.46 \\
                   & Classical & Yes & 96.62 ± 20.57 & 112.16 \\
\hline
\multirow{2}{*}{5} & Quantum   & Yes & 90.29 ± 3.61  & 118.56 \\
                   & Classical & Yes & 93.19 ± 4.46  & 117.11 \\
\hline
\end{tabular}
\label{tab:mid_cantilever_compliance_diversity}
\end{table}

\bibliographystyle{elsarticle-num}
\bibliography{bibliography.bib}

\section*{Appendix A. Step-by-step derivation of the quantum state and expectation values}

We consider a two-qubit variational quantum circuit with one repetition layer, comprising two layers of parameterized single-qubit \( R_Y \) rotations and an intermediate CNOT entangling gate. The goal of this appendix is to explicitly derive the resulting quantum state from this circuit and to compute the expectation values of single-qubit Pauli observables \( X_i, Y_i, Z_i \) acting on each qubit. 

To illustrate the process in full detail, we assume specific values for the variational parameters \( \boldsymbol{\theta} \), allowing us to perform all matrix multiplications explicitly and obtain numerical results. The subscript \( \theta_{r,i} \) indicates the rotation angle applied in repetition layer \( r \in \{0,1\} \) on qubit \( i \in \{0,1\} \).

The overall unitary transformation implemented by the circuit is given by
\[
U(\boldsymbol{\theta}) = \left( R_Y(\theta_{1,0}) \otimes R_Y(\theta_{1,1}) \right) \cdot \text{CNOT}_{0 \to 1} \cdot \left( R_Y(\theta_{0,0}) \otimes R_Y(\theta_{0,1}) \right),
\]
where each \( R_Y(\theta) \) gate is defined as
\[
R_Y(\theta) =
\begin{bmatrix}
\cos(\theta/2) & -\sin(\theta/2) \\
\sin(\theta/2) & \cos(\theta/2)
\end{bmatrix}.
\]
The tensor product \( R_Y(\theta_0) \otimes R_Y(\theta_1) \) expands to
\[
\begin{aligned}
R_Y(\theta_0) \otimes R_Y(\theta_1) =
\begin{bmatrix}
\cos(\tfrac{\theta_0}{2}) \cos(\tfrac{\theta_1}{2}) &
-\cos(\tfrac{\theta_0}{2}) \sin(\tfrac{\theta_1}{2}) &
-\sin(\tfrac{\theta_0}{2}) \cos(\tfrac{\theta_1}{2}) &
\sin(\tfrac{\theta_0}{2}) \sin(\tfrac{\theta_1}{2}) \\
\cos(\tfrac{\theta_0}{2}) \sin(\tfrac{\theta_1}{2}) &
\cos(\tfrac{\theta_0}{2}) \cos(\tfrac{\theta_1}{2}) &
-\sin(\tfrac{\theta_0}{2}) \sin(\tfrac{\theta_1}{2}) &
-\sin(\tfrac{\theta_0}{2}) \cos(\tfrac{\theta_1}{2}) \\
\sin(\tfrac{\theta_0}{2}) \cos(\tfrac{\theta_1}{2}) &
-\sin(\tfrac{\theta_0}{2}) \sin(\tfrac{\theta_1}{2}) &
\cos(\tfrac{\theta_0}{2}) \cos(\tfrac{\theta_1}{2}) &
-\cos(\tfrac{\theta_0}{2}) \sin(\tfrac{\theta_1}{2}) \\
\sin(\tfrac{\theta_0}{2}) \sin(\tfrac{\theta_1}{2}) &
\sin(\tfrac{\theta_0}{2}) \cos(\tfrac{\theta_1}{2}) &
\cos(\tfrac{\theta_0}{2}) \sin(\tfrac{\theta_1}{2}) &
\cos(\tfrac{\theta_0}{2}) \cos(\tfrac{\theta_1}{2})
\end{bmatrix}.
\end{aligned}
\]

\subsection*{A.1. Initial state and first rotation layer}

We begin with the initial state
\[
|\psi_0\rangle = |00\rangle = 
\begin{bmatrix}
1 & 0 & 0 & 0
\end{bmatrix}^\top.
\]

The first layer applies single-qubit \( R_Y \) rotations with
\[
\theta_{0,0} = \frac{\pi}{3}, \quad \theta_{0,1} = \frac{\pi}{4}.
\]

The corresponding numerical matrices are
\[
R_Y\left(\tfrac{\pi}{3}\right) \approx 
\begin{bmatrix}
0.8660 & -0.5000 \\
0.5000 & 0.8660
\end{bmatrix}, \quad
R_Y\left(\tfrac{\pi}{4}\right) \approx 
\begin{bmatrix}
0.9239 & -0.3827 \\
0.3827 & 0.9239
\end{bmatrix}.
\]

The full rotation operator is
\[
U_1 = R_Y(\theta_{0,0}) \otimes R_Y(\theta_{0,1}) \approx
\begin{bmatrix}
0.8001 & -0.3314 & -0.4619 & 0.1913 \\
0.3308 & 0.8001 & -0.1913 & -0.4619 \\
0.4619 & -0.1913 & 0.8001 & -0.3314 \\
0.1913 & 0.4619 & 0.3308 & 0.8001
\end{bmatrix}.
\]

Multiplying by the initial state gives
\[
|\psi_1\rangle = U_1 |\psi_0\rangle \approx 
\begin{bmatrix}
0.8001 & 0.3308 & 0.4619 & 0.1913
\end{bmatrix}^\top.
\]

\subsection*{A.2. Entangling layer (CNOT gate)}

We apply a CNOT gate (control: qubit 0, target: qubit 1)
\[
\text{CNOT} =
\begin{bmatrix}
1 & 0 & 0 & 0 \\
0 & 1 & 0 & 0 \\
0 & 0 & 0 & 1 \\
0 & 0 & 1 & 0
\end{bmatrix}, \quad
|\psi_2\rangle = \text{CNOT} \cdot |\psi_1\rangle \approx
\begin{bmatrix}
0.8001 & 0.3308 & 0.1913 & 0.4619
\end{bmatrix}^\top.
\]

\subsection*{A.3. Second rotation layer}

We now apply a second set of \( R_Y \) rotations with
\[
\theta_{1,0} = \frac{\pi}{6}, \quad \theta_{1,1} = \frac{\pi}{5},
\]
and matrices:
\[
R_Y\left(\tfrac{\pi}{6}\right) \approx 
\begin{bmatrix}
0.9659 & -0.2588 \\
0.2588 & 0.9659
\end{bmatrix}, \quad
R_Y\left(\tfrac{\pi}{5}\right) \approx 
\begin{bmatrix}
0.9511 & -0.3090 \\
0.3090 & 0.9511
\end{bmatrix}.
\]

The resulting two-qubit operator is
\[
U_2 = R_Y(\theta_{1,0}) \otimes R_Y(\theta_{1,1}) \approx
\begin{bmatrix}
0.9192 & -0.2982 & -0.2468 & 0.0775 \\
0.2886 & 0.9203 & -0.0775 & -0.2468 \\
0.2468 & -0.0775 & 0.9192 & -0.2982 \\
0.0775 & 0.2468 & 0.2886 & 0.9203
\end{bmatrix}.
\]

The final quantum state is
\[
|\psi(\boldsymbol{\theta})\rangle = U_2 \cdot |\psi_2\rangle \approx
\begin{bmatrix}
0.6259 & 0.4143 & 0.2083 & 0.6270
\end{bmatrix}^\top.
\]

\subsection*{A.4. Expectation values of single-qubit Pauli observables}

Let \( P \in \{X, Y, Z\} \). For each qubit \( i \in \{0,1\} \), we define the corresponding Pauli observable acting on the full Hilbert space as
\[
P_i = I^{\otimes i} \otimes P \otimes I^{\otimes(n - i - 1)}.
\]
We compute the expectation values:
\[
z_i = \langle \psi | Z_i | \psi \rangle, \quad
x_i = \langle \psi | X_i | \psi \rangle, \quad
y_i = \langle \psi | Y_i | \psi \rangle.
\]

For the real-valued state
\[
|\psi\rangle \approx 
\begin{bmatrix}
\psi_0 & \psi_1 & \psi_2 & \psi_3
\end{bmatrix}^\top =
\begin{bmatrix}
0.6259 & 0.4143 & 0.2083 & 0.6270
\end{bmatrix}^\top,
\]
we obtain

\begin{align*}
z_0 &= |\psi_0|^2 + |\psi_1|^2 - |\psi_2|^2 - |\psi_3|^2 \approx 0.3914, \\
z_1 &= |\psi_0|^2 - |\psi_1|^2 + |\psi_2|^2 - |\psi_3|^2 \approx -0.0773, \\
x_0 &= 2(\psi_0 \psi_2 + \psi_1 \psi_3) \approx 0.7799, \\
x_1 &= 2(\psi_0 \psi_1 + \psi_2 \psi_3) \approx 0.7802, \\
y_0 &= 2\,\mathrm{Im}(-\psi_0 \psi_3 + \psi_1 \psi_2) = 0 \quad (\text{real state}), \\
y_1 &= 2\,\mathrm{Im}(-\psi_0 \psi_1 + \psi_2 \psi_3) = 0 \quad (\text{real state}).
\end{align*}

The final results are summarized below

\begin{center}
\begin{tabular}{|c|c|c|c|}
\hline
Qubit \( i \) & \( z_i \) & \( x_i \) & \( y_i \) \\
\hline
0 & 0.3914 & 0.7799 & 0 \\
1 & –0.0773 & 0.7802 & 0 \\
\hline
\end{tabular}
\end{center}

These expectation values represent measurable observables in the Pauli basis and can be used as features in variational quantum eigensolvers or hybrid quantum-classical optimization pipelines.

\end{document}